\documentclass[longauth]{aa}
\usepackage{color}
\usepackage{graphicx}
\usepackage{txfonts}

\newcommand{\ms}{m\,s$^{-1}$}
\newcommand {\logRHK}{\ensuremath{\log R'_\mathrm{HK}}}

\def\vsini{\ensuremath{{\upsilon}\sin i}}
\def\kms{$\mathrm{km\,s}^{-1}$}
\def\ms{\hbox{\,m\,s$^{-1}$}}         
\def\m2s2{\hbox{\,m$^{2}$\,s$^{-2}$}} 
\def\kms{\hbox{\,km\,s$^{-1}$}}       
\def\vsini{\hbox{$v$\,sin\,$i$}}      

\def\rs{R$_{\rm s}$}

\def\rp{R_{\rm p}}

\begin{document} 

      \title{Atmospheric Rossiter-McLaughlin effect and transmission spectroscopy of WASP-121b with ESPRESSO\thanks{Based in part on Guaranteed Time Observations collected at the European Southern Observatory under ESO programme 1102.C-0744 by the ESPRESSO Consortium.}}
   \titlerunning{}
   \authorrunning{F.~Borsa et al.}
\author{F.~Borsa\inst{\ref{brera}}
                \and R.~Allart\inst{\ref{Geneve}}
                \and N.~Casasayas-Barris\inst{\ref{IAC},\ref{IAC2}}
                \and H.~Tabernero\inst{\ref{porto}}
                \and M.~R.~Zapatero~Osorio\inst{\ref{csic-inta}}
                \and S.~Cristiani\inst{\ref{trieste},\ref{cristiani}}
                \and F.~Pepe\inst{\ref{Geneve}}
                \and R.~Rebolo\inst{\ref{IAC},\ref{IAC2},\ref{rebolo}}
                                \and N.~C.~Santos\inst{\ref{porto},\ref{porto2}}                                
                        \and V.~Adibekyan\inst{\ref{porto},\ref{porto2}}
                                 \and V.~Bourrier\inst{\ref{Geneve}}
                                \and O.~D.~S.~Demangeon\inst{\ref{porto},\ref{porto2}}
                                \and D.~Ehrenreich\inst{\ref{Geneve}}
                          \and E.~Pall\'e \inst{\ref{IAC},\ref{IAC2}}   
                                \and S.~Sousa\inst{\ref{porto}}
                                 \and J.~Lillo-Box\inst{\ref{csic-inta}}        
                                \and C.~Lovis\inst{\ref{Geneve}}
                                \and G.~Micela\inst{\ref{palermo}}
                                \and M.~Oshagh\inst{\ref{IAC},\ref{IAC2}}
                                \and E.~Poretti\inst{\ref{brera},\ref{tng}} %
                                        \and A.~Sozzetti\inst{\ref{oato}}       
                \and C.~Allende~Prieto\inst{\ref{IAC},\ref{IAC2}}%
                \and Y.~Alibert\inst{\ref{berna}}%
                \and M.~Amate\inst{\ref{IAC},\ref{IAC2}}
                \and W.~Benz\inst{\ref{berna}}
                \and F.~Bouchy\inst{\ref{Geneve}}%
                \and A.~Cabral\inst{\ref{lisbona},\ref{lisbona2}}       
                \and H.~Dekker \inst{\ref{garching}}
                \and V.~D'Odorico\inst{\ref{trieste},\ref{cristiani}}%
                \and P.~Di Marcantonio\inst{\ref{trieste}} %
                \and P.~Figueira\inst{\ref{esocile},\ref{porto}}%
                \and R.~Genova~Santos\inst{\ref{IAC},\ref{IAC2}}%
                \and J.~I.~Gonz\'alez~Hern\'andez\inst{\ref{IAC},\ref{IAC2}}%
                \and G.~Lo~Curto\inst{\ref{esocile}}%
                \and A.~Manescau\inst{\ref{garching}}
                \and C.~J.~A.~P.~Martins\inst{\ref{porto},\ref{porto3}}%
                \and D.~M\'egevand\inst{\ref{Geneve}}
                \and A.~Mehner\inst{\ref{esocile}}%
                \and P.~Molaro\inst{\ref{trieste},\ref{cristiani}}%
                \and N.~Nunes\inst{\ref{lisbona},\ref{lisbona2}}%
                \and M.~Riva\inst{\ref{brera}}%
                \and A.~Su\'arez Mascare\~{n}o\inst{\ref{IAC},\ref{IAC2}}%
                \and S.~Udry\inst{\ref{Geneve}}%
                \and F.~Zerbi\inst{\ref{brera}}
                }                       
 \institute{INAF -- Osservatorio Astronomico di Brera, Via E. Bianchi 46, 23807 Merate (LC), Italy \label{brera}
        \and D\'epartement d'astronomie, Universit\'e de Gen\`eve, ch. des Maillettes 51, 1290 Versoix, Switzerland \label{Geneve}
        \and INAF -- Osservatorio Astrofisico di Torino, Via Osservatorio 20, I-10025 Pino Torinese, Italy \label{oato}
        \and Fundaci\'on G. Galilei - INAF (TNG), Rambla J. A. Fern\'andez P\'erez 7, E-38712 Bre\~{n}a Baja (La Palma), Spain \label{tng}
          \and INAF -- Osservatorio Astronomico di Trieste, via Tiepolo 11, 34143 Trieste, Italy \label{trieste}
        \and Instituto de Astrof\'{\i}sica de Canarias, C/V\'{\i}a L\'actea s/n, E-38205 La Laguna (Tenerife), Spain \label{IAC} 
        \and Universidad de La Laguna (ULL), Departamento de Astrof\'{\i}sica, 38206 La Laguna, Tenerife, Spain  \label{IAC2} 
\and Centro de Astrobiolog\'{\i}a (CSIC-INTA), Carretera de Ajalvir km 4 - 28850 Torrej\'{o}n de Ardoz, Madrid, Spain \label{csic-inta}
 \and INAF -- Osservatorio Astronomico di Palermo, Piazza del Parlamento, 1, 90134, Palermo, Italy \label{palermo}
 \and Instituto de Astrof\'{\i}sica e Ci\^{e}n\c{c}ias do Espa\c{c}o, Universidade do Porto, CAUP, Rua das Estrelas, 4150-762 Porto, Portugal \label{porto}
 \and Departamento de F\'{\i}sica e Astronomia, Faculdade de Cien\c{c}ias, Universidade do Porto, Rua do Campo Alegre, 4169-007 Porto, Portugal\label{porto2}
 \and Centro de Astrof\'{\i}sica da Universidade do Porto, Rua das Estrelas, 4150-762 Porto, Portugal\label{porto3}
 \and European Southern Observatory, Karl-Schwarzschild-Strasse 2, 85748 Garching b. Munchen, Germany\label{garching}
 \and Universitat Bern, Physikalisches Institut, Siedlerstrasse 5, 3012 Bern, Switzerland \label{berna}
 \and Faculdade de Ci\^{e}n\c{c}ias da Universidade de Lisboa (Departamento de F\'{\i}sica), Edif\'{\i}cio C8, 1749-016 Lisboa, Portugal \label{lisbona}
\and Instituto de Astrof\'{\i}sica e Ci\^{e}n\c{c}ias do Espa\c{c}o, Universidade de Lisboa, Edif\'{\i}cio C8, 1749-016 Lisboa, Portugal\label{lisbona2}
\and Consejo Superior de Investigaciones Cient\'ificas, E-28006 Madrid, Spain \label{rebolo}
\and ESO, European Southern Observatory, Alonso de Cordova 3107, Vitacura, Santiago\label{esocile}
\and Institute for Fundamental Physics of the Universe, IFPU, Via Beirut 2, 34151 Grignano, Trieste, Italy\label{cristiani}
}
             \offprints{F.~Borsa\\ \email{francesco.borsa@inaf.it}}

   \date{Received ; accepted }

 
  \abstract
   {Ultra-hot Jupiters are excellent laboratories for the study of exoplanetary atmospheres. \object{WASP-121b} is one of the most studied; many recent analyses of its atmosphere report interesting features at different wavelength ranges.}
   {In this paper we analyze one transit of WASP-121b acquired with the high-resolution spectrograph ESPRESSO at VLT in one-telescope mode, and one partial transit taken during the commissioning of the instrument in four-telescope mode.  
   }
   {
   We take advantage of the very high S/N data and of the extreme stability of the spectrograph to investigate the anomalous in-transit radial velocity curve and study the transmission spectrum of the planet. 
   We pay particular attention to the removal of instrumental effects, and stellar and telluric contamination. The transmission spectrum is investigated through single-line absorption and cross-correlation with theoretical model templates. 
   }
   {By analyzing the in-transit radial velocities we were able to infer the presence of the atmospheric Rossiter-McLaughlin effect. We measured the height of the planetary atmospheric layer that correlates with the stellar mask (mainly Fe) to be 1.052$\pm$0.015 $\rp$ and we also confirmed the blueshift of the planetary atmosphere. 
  By examining the planetary absorption signal on the stellar cross-correlation functions we confirmed the presence of a temporal variation of its blueshift during transit, which could be investigated spectrum-by-spectrum thanks to the quality of our ESPRESSO data. 
We detected significant absorption in the transmission spectrum for Na, H, K, Li, \ion{Ca}{ii}, and Mg, and we certified their planetary nature by using the 2D tomographic technique. 
Particularly remarkable is the detection of Li, with a line contrast of $\sim$0.2\% detected at the 6$\sigma$ level. 
With the cross-correlation technique we confirmed the presence of \ion{Fe}{i}, \ion{Fe}{ii}, \ion{Cr}{i}, and \ion{V}{i}.
H$\alpha$ and \ion{Ca}{ii} are present up to very high altitudes in the atmosphere ($\sim$1.44 $\rp$ and $\sim$2 $\rp$, respectively), and also extend beyond the transit-equivalent Roche lobe radius of the planet. 
These layers of the atmosphere have a large line broadening that is not compatible with being caused by the tidally locked rotation of the planet alone, and could arise from vertical winds or high-altitude jets in the evaporating atmosphere. 
}
   {}

   \keywords{planetary systems --  techniques: spectroscopic  -- techniques: radial velocities  -- planets and satellites: atmospheres -- stars:individual:WASP-121}

   \maketitle
%

\section{Introduction\label{sec:intro}}

Ultra-hot Jupiters (UHJs) are giant exoplanets on short-period orbits (P$\le$3 days) that receive very intense irradiation from their host stars (equilibrium temperature T$_{\rm eq}\ge$2000 K). As a consequence, their expanded atmospheres are in extreme states, experiencing phenomena like atmospheric evaporation and escape \citep[e.g.,][]{2003Natur.422..143V,2010ApJ...714L.222F,2018NatAs...2..714Y,2018haex.bookE.148B,sing}.
Most of the major molecules (except CO) are thermally dissociated into their atomic constituents in the atmospheres of UHJs \citep{2018A&A...617A.110P,2018ApJ...866...27L}. 
Given their high equilibrium temperatures, iron is also often found in the gaseous state \citep[e.g.,][]{2018Natur.560..453H,2019A&A...628A...9C,espresso1,cabot}.

High-resolution transmission spectroscopy is one of the most powerful tools to study exoplanetary atmospheres \citep[e.g.,][]{2017JGRE..122...53D}. While investigating different layers of the planetary atmospheres, it resolves the possible ambiguities present in low-resolution spectra \citep[e.g., molecular identification and abundance determinations when multiple species overlap;][]{2017ApJ...839L...2B}, and it allows us to  spectroscopically separate the planetary and stellar restframes \citep[e.g.,][]{2010Natur.465.1049S,2016ApJ...817..106B,2018A&A...617A.134B,2019A&A...628A...9C}. Recently,  in-transit radial velocities (RVs) also demonstrated their capacity to scan the atmosphere of hot exoplanets. In-transit RVs are affected by a deviation from the Keplerian motion around the star--planet center of mass, known as the Rossiter-McLaughlin effect \citep[RM;][]{1924ApJ....60...15R,1924ApJ....60...22M}, which is caused by the stellar rotation. The shape and amplitude of this deformation depend on the projected spin-orbit angle \citep[e.g.,][]{2008ApJ...682.1283W}, on \vsini\ 
and the radius of the planet at any given wavelength \citep{2004MNRAS.353L...1S,2009A&A...499..615D,2015A&A...580A..84D}. In addition, if the atmosphere of the planet contains elements that are present in the stellar mask used to determine the RVs (e.g., Fe), another deformation is added to the classic RM, whose amplitude is proportional to the height of the atmospheric layers that correlate with the stellar mask \citep[the atmospheric RM;][]{atmoRM}. 

WASP-121b \citep[$M_{\rm p}\sim1.2$ M$_{\rm J}$, $R_{\rm p}\sim1.7$ R$_{\rm J}$,][]{delrez} is one of the most studied transiting UHJs. 
Orbiting a bright (V=10.4 mag) F6 star in $\sim$1.27 days and with a misaligned projected spin-orbit angle \citep[$\lambda=87.2^\circ$,][]{bourrier}, it is an excellent target for atmospheric follow-up investigation.
\citet{2019A&A...623A..57S} found excess absorption in near-UV transit observations with the {\it Swift} satellite, possibly caused by \ion{Fe}{ii} in a dense extended atmosphere.
\ion{Fe}{ii} was then confirmed in its exosphere with {\it Hubble Space Telescope} ({\it HST}) observations, as was \ion{Mg}{ii} \citep{sing}.
{\it HST} also provided an optical transmission \citep{2018AJ....156..283E} and optical+infrared emission spectra, which led to the detection of water in emission \citep{2017Natur.548...58E} and to an estimation of the metallicity and C/O  \citep{2019MNRAS.488.2222M}.
A secondary eclipse was also   detected with a 1m class telescope at 2$\mu m$ \citep{2019A&A...625A..80K}.
Phase curve observations with TESS pointed to a temperature inversion possibly caused by VO and TiO \citep{bourrier_phasecurve}, but there was no evidence of these compounds at the terminator \citep{2020arXiv200202795M}. 
With high-resolution UVES transit observations, \citet{gibson} detected \ion{Fe}{i} low in the atmosphere.
Using HARPS transit observations, \citet{bourrier}  again found evidence for the presence of iron in the atmosphere by discovering that it correlates with 
the stellar mask used to extract the cross-correlation functions (CCFs). With the same dataset \citet{cabot} detected \ion{Na}{i}, H$\alpha$, and again \ion{Fe}{i}, and \citet{2020arXiv200605995B}  also found evidence of the presence of \ion{Cr}{i} and \ion{V}{i}.

In this manuscript we present the analysis of high-resolution spectroscopic transit observations of WASP-121b with the Echelle Spectrograph for Rocky Exoplanets and Stable Spectroscopic Observations \citep[ESPRESSO;][]{2010SPIE.7735E..0FP,espresso,espresso2020}.
We first present the observations (Sect. \ref{sec:data_sample}). Then we perform a stellar characterization (Sect. \ref{sec:stellar_par}), followed by an analysis of the in-transit RVs and of the planetary atmospheric signal imprinted in the stellar cross-correlation functions (Sect. \ref{sec:rv}). We then present the analysis of the planetary transmission spectrum (Sect. \ref{sec:trans_spectrum}), and finally discuss our results (Sect. \ref{sec:discussion}).


\begin{table*}[!ht]
\begin{center}
\caption{WASP-121 ESPRESSO observations log.} 
\label{tab:log}
\footnotesize
\begin{tabular}{ccccccccc}
 \hline\hline
 \noalign{\smallskip}
Night & Exp time & Mode & fiber B & S/N$^{1}$@550nm & Time on target [hours] &N$_{\rm spectra}$&Seeing$_{\rm ave}$ [arcsec]\\ 
 \noalign{\smallskip}
 \hline
\noalign{\smallskip}
06 Jan 2019  & 400 s & 1-UT HR21&Sky & $\sim$45 & 6.3 & 52 & 0.83\\
30 Nov 2018  & 300 s & 4-UT MR42&Fabry-Perot & $\sim$180 & 2.7 & 29 & 0.79\\
\noalign{\smallskip}
 \hline
\noalign{\smallskip}
 \multicolumn{5}{l}{$^{1}$\footnotesize{per pixel extracted.}} \\
\end{tabular}
\end{center}
\end{table*}

\section{Data sample\label{sec:data_sample}}

We analyze two transits of WASP-121b observed with ESPRESSO. ESPRESSO is a fiber-fed, ultra-stabilized echelle high-resolution spectrograph, with the capability of collecting light from each 8.2m Unit Telescope (UT) of the Very Large Telescope (VLT) individually or the four UTs simultaneously, yielding a 16m-equivalent telescope. 
It is the first instrument with this capability, and  it is located in the VLT Combined-Coudé Laboratory.
ESPRESSO covers the optical wavelength range 3800-7880 $\AA$.

One transit of WASP-121b was observed as part of the Guaranteed Time Observation in 1-UT HR21 mode (resolving power R$\sim$138,000) under program 1102.C-0744, using UT3 Melipal. 
We also analyze one partial transit of WASP-121b observed in  4-UT MR42 mode (R$\sim$70,000) during the commissioning of the instrument. Despite the lower resolution with respect to the HR mode, the 4-UT MR spectra benefit from the higher signal-to-noise ratio (S/N). Observing conditions were good and comparable during the two transits, with an average seeing of $\sim$0.8.
A log of the observations is presented in Table \ref{tab:log}.

For the 1-UT transit, while   fiber A of the instrument was observing the target, fiber B was pointing to the sky.
Instead, for the 4-UT transit fiber B was put on the Fabry-Perot simultaneous reference.
We analyzed the 1D spectra  extracted by the Data Reduction Software (DRS) pipeline\footnote{publicly available at www.eso.org/sci/software/pipelines/espresso/espresso-pipe-recipes.html}. 
In the case of the 1-UT transit, we analyzed the sky-subtracted 1D spectra.


\section{Stellar characterization\label{sec:stellar_par}}

We calculated the stellar parameters of WASP-121 by means of a spectral synthesis analysis on a combined spectrum of the 1-UT night using the SteParSyn code \citep{2018MNRAS.476.3106T}.
It uses a Markov chain Monte Carlo process to derive the probability distribution functions of the stellar atmospheric parameters through a grid of synthetic spectra, 
using MARCS models \citep{2008A&A...486..951G} and the \ion{Fe}{i}-\ion{Fe}{ii} line list for metal-rich dwarf stars described in \citet{2019A&A...628A.131T}.
The derived T$_{\rm eff}$, log g, and metallicity (Table \ref{tabParameters}) are in agreement with the values of \citet{delrez}, while the \vsini\ value calculated with line broadening is slightly lower than theirs ($11.8\pm0.2$ versus $13.5\pm0.7$ \kms).
Using the Padova stellar model isochrones\footnote{http://stev.oapd.inaf.it/cgi-bin/param\_1.3} and the Gaia DR2 parallax \citep[$3.676\pm0.021$ mas;][]{2018yCat.1345....0G}, we obtain a stellar age of $1.03\pm0.43$ Gyr, a stellar mass of $1.38\pm0.02$ M$_{\sun}$, and a stellar radius of $1.44\pm0.03$ R$_{\sun}$, also consistent with the values reported by \citet{delrez}.
The results are shown in Table \ref{tabParameters}.

\begin{table}
\begin{center}
\caption{Properties of the WASP-121 system.}
\label{tabParameters}
\footnotesize
\begin{tabular}{ccc}
 \hline\hline
 \noalign{\smallskip}
 Parameter & Value & Reference\\
 \noalign{\smallskip}
 \hline
\noalign{\smallskip}
\multicolumn{3}{c}{WASP-121}\\
\noalign{\smallskip}
\hline
\noalign{\smallskip}
\multicolumn{3}{c}{\it Stellar parameters}\\
\noalign{\smallskip}
\noalign{\smallskip}
T [K] &   6586 $\pm$ 59 & This work\\
log g &  4.47 $\pm$ 0.08& This work\\
 Fe/H & 0.13  $\pm$ 0.04& This work\\
\vsini \ [\kms] &  11.8 $\pm$ 0.2& This work\\
$M_{\rm s}$ [M$_{\sun}$] &$1.38\pm0.02$ &  This work\\
$R_{\rm s}$ [R$_{\sun}$] &$1.44\pm0.03$ &  This work\\
age [Gyr] & $1.03\pm0.43$&  This work\\
\noalign{\smallskip}
\noalign{\smallskip}
\multicolumn{3}{c}{\it Orbital parameters}\\
\noalign{\smallskip}
\noalign{\smallskip}
Period [days]&   $1.27492504_{-1.4\times10^{-7}}^{+1.5\times10^{-7}}$ & \citet{bourrier}\\
$T_0$ [BJD] &   $2458119.72074 \pm 0.00017$ & \citet{bourrier_phasecurve}\\  
$R_{\rm p}$/$R_{\rm s}$ &   0.12534$\pm$0.00005 &  \citet{bourrier}\\
$a/R_{\rm s}$ &   $3.8131_{-0.0060}^{+0.0075}$ & \citet{bourrier}\\
$e$ &   $0.0$ & assumed\\
$i$ [degrees]&   $88.49\pm 0.16$ & \citet{bourrier}\\
V$_{\rm sys}$ [\kms] & $38.198\pm 0.002$ & This work\\
$M_{\rm p}$ [M$_{\rm Jup}$] &$1.157\pm0.070$&  \citet{bourrier}\\
$K_{\rm s}$ [\kms] & 0.177$\pm$0.008& \citet{bourrier}\\
\noalign{\smallskip}
\noalign{\smallskip}
\multicolumn{3}{c}{\it RM+RM$_{atmo}$ fit}\\
\noalign{\smallskip}
\noalign{\smallskip}
 $\mu$ &  $0.68_{-0.08}^{+0.09}$  & This work\\
$ \lambda$ [degrees]&  $-87.08_{-0.27}^{+0.29}$  & This work\\
\vsini\ [\kms]&  15.4$\pm$0.8  & This work\\
$R_{\rm p,atmo}$/$R_{\rm s}$ &  $-0.044_{-0.004}^{+0.003}$  &  This work\\
delay [phase]&  $0.0029_{-0.0008}^{+0.0005}$  & This work\\
$a/R_{\rm s}$ $ratio$ &  $2.61_{-0.13}^{+0.21}$ & This work\\
\noalign{\smallskip}
 \hline
\end{tabular}
\end{center}
\end{table}

\subsection{Fourier transform of the CCF\label{sec:fourier}}

Because of the moderate value of \vsini=11.8\kms\ found with the spectral analysis,
we also calculated it by means of the Fourier transform of the mean line profile \citep[e.g.,][]{dravins1990}. This method can be used when the stellar \vsini\ is much greater than the width of the instrumental profile, typically for stars with \vsini\ $\gtrsim$10 \kms\ \citep{2002A&A...384..155R}. We used the CCF as mean line profile since it has been proven to be a good mean line profile indicator for this kind of study \citep{tauboo}.
The $q$ positions of the first zeros of the Fourier transform give an estimate of the projected rotational velocity \vsini, while the ratio $q_{2}/q_{1}$ is a direct indicator for solar-like differential rotation \citep[equator rotating faster than poles;][]{2002A&A...384..155R}.
We found \vsini $= 11.90 \pm 0.31$ \kms\ and $q_{2}/q_{1} = 1.43 \pm 0.28$.
By adopting an inclination angle for the star $i_\star = 8.1^\circ$ \citep{bourrier} and using the $q_{2}/q_{1}$ value, we computed the differential rotation parameter $\alpha$ by following \citet{2003A&A...398..647R}.
We found $\alpha = 0.06 \pm 0.03$, compatible with the value $\alpha=0.08^{+0.11}_{-0.13}$ obtained by \citet{bourrier} with the study of the reloaded RM effect, thus excluding high differential rotation rates.
We note that differential rotation can contribute to the RM shape \citep[e.g.,][]{2020MNRAS.493.5928S}, in particular for polar orbit planets like WASP-121b. However, given its low rate for WASP-121, it is unlikely that it plays a major role in this context, but it could explain some of the RV fit residuals (Sect. \ref{sec.atmoRM}).

\subsection{Stellar activity\label{sec:activity}}

\begin{figure}[h]
\centering
\includegraphics[width=\linewidth]{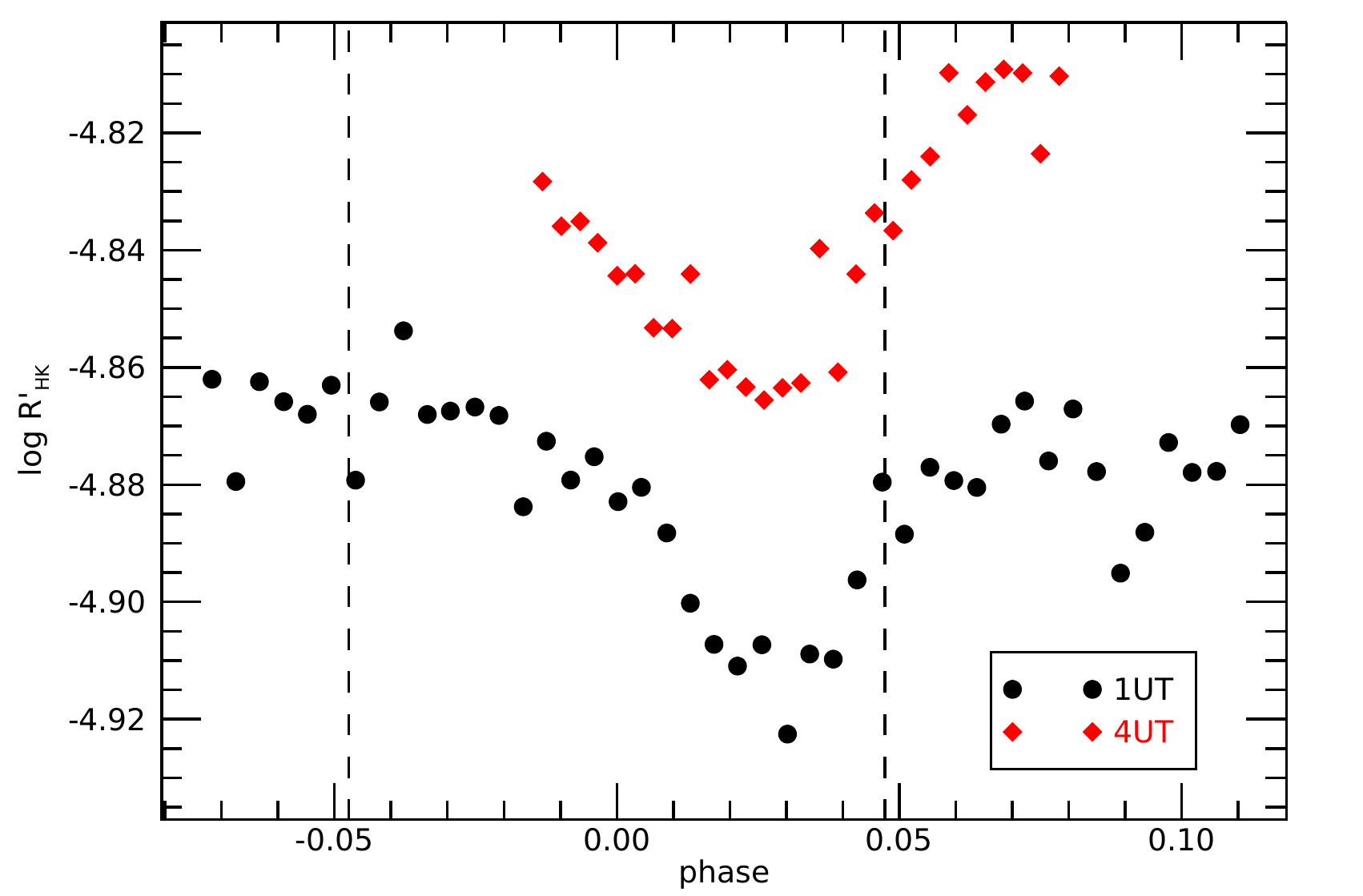}
\caption{\logRHK \ activity indices calculated for the two transits. Vertical dotted lines show the expected beginning and end of the transit. Phase 0 represents the mid-transit time.}
\label{fig.rhk}
\end{figure}

Given the difference in the RV slope (see Sect. \ref{sec:rv}) and the possible different line contrast observed in the planetary absorptions of the two transits (see Sect. \ref{sec:detections}), we calculated the level of stellar activity \logRHK\ \citep{1984ApJ...279..763N} for each transit, using the HARPS index calibration of \citet{astu17} as the one for ESPRESSO is still not available \citep{2020A&A...635A..13F}. 
Since there is a strong in-transit planetary signal in the core of the \ion{Ca}{ii} H\&K lines (Fig. \ref{fig.rhk}, see also Sect. \ref{sec:detections}), for each transit we average the \logRHK\ values calculated on the out-of-transit spectra only.  We find \logRHK\ =-4.87 $\pm$ 0.01 and \logRHK\ =-4.81 $\pm$ 0.01 for the 1-UT and 4-UT transits, respectively. The star is thus more active during the partial transit observed with the 4-UT mode. 
We note that stellar activity can change RM shape and/or out of transit RV slope significantly \citep{2018A&A...619A.150O}.

The in-transit behavior  of the \logRHK\ index (Fig. \ref{fig.rhk}), which shows the same pattern in both transits, is likely due to the passage of the planet combined with atmospheric absorption in the \ion{Ca}{ii} H\&K lines (Sect. \ref{sec:detections}). 
It is however interesting to note one thing: 
the rotation period of the star is $\sim$1.13 days \citep{delrez,bourrier}, derived from spectroscopy, while TESS photometry showed a similar periodicity at $\sim$1.16 days \citep{bourrier}. 
If the stellar rotation period were 1.1554 days, then the star would have accomplished exactly 32 rotations between the two observations, and thus the planet could also be blocking the same active region on the stellar surface at the two transits.


\begin{figure}
\centering
\includegraphics[width=\linewidth]{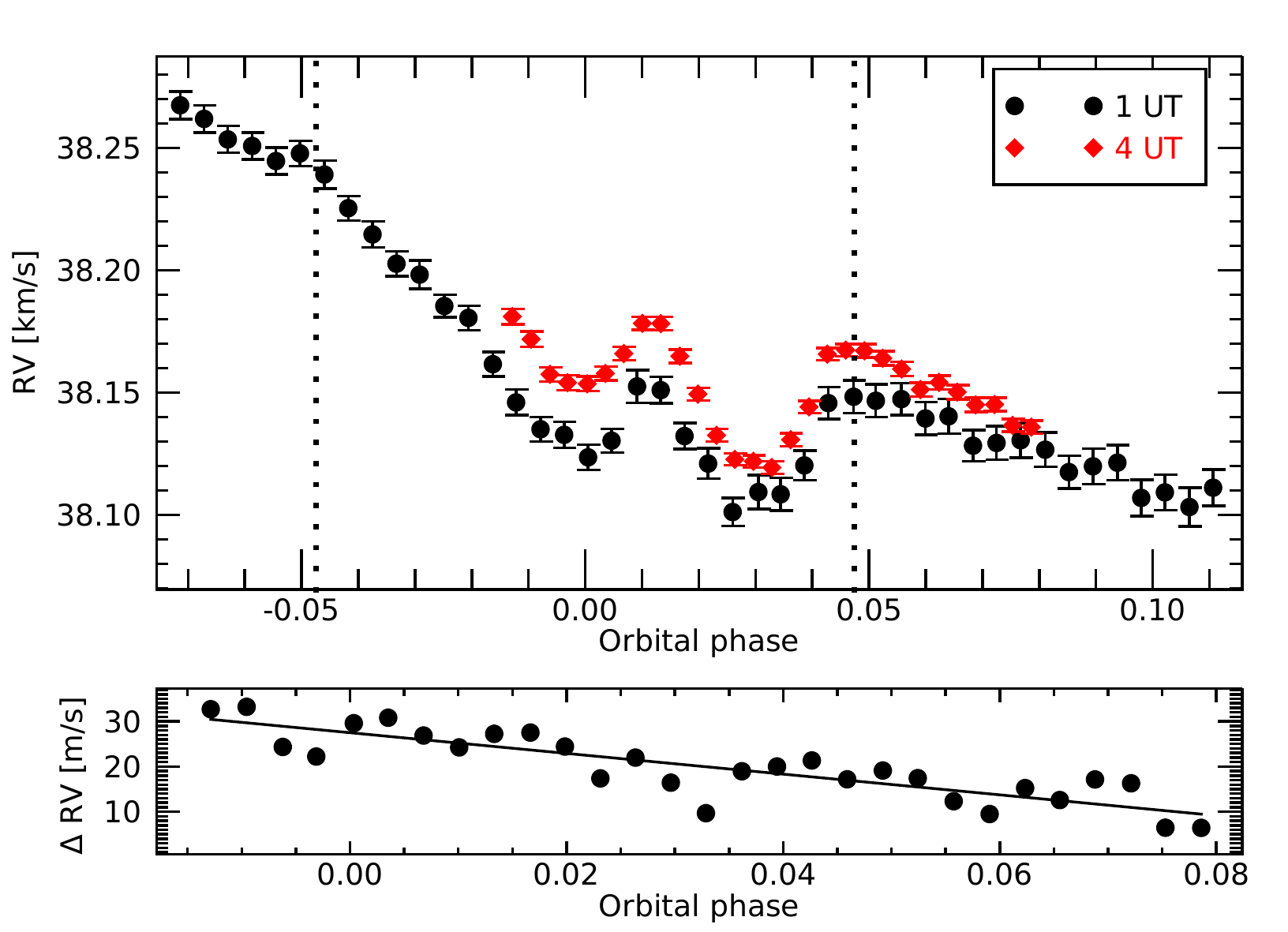}
\caption{{\it (Top panel)} ESPRESSO radial velocities of the WASP-121b transits observed. Vertical dotted lines indicate the expected beginning and end of the transit. {\it (Bottom panel)} RV difference between the two transits, calculated by quadratically interpolating the 1-UT RVs on the 4-UT phases. The black line shows the fitted linear trend with a slope significant at the 4.5$\sigma$ level. An offset is expected because of the different instrumental setups used in the two transits.}
\label{fig.rv}
\end{figure}

\section{Radial velocities and CCFs analysis \label{sec:rv}}

\begin{figure}
\centering
\includegraphics[width=\linewidth]{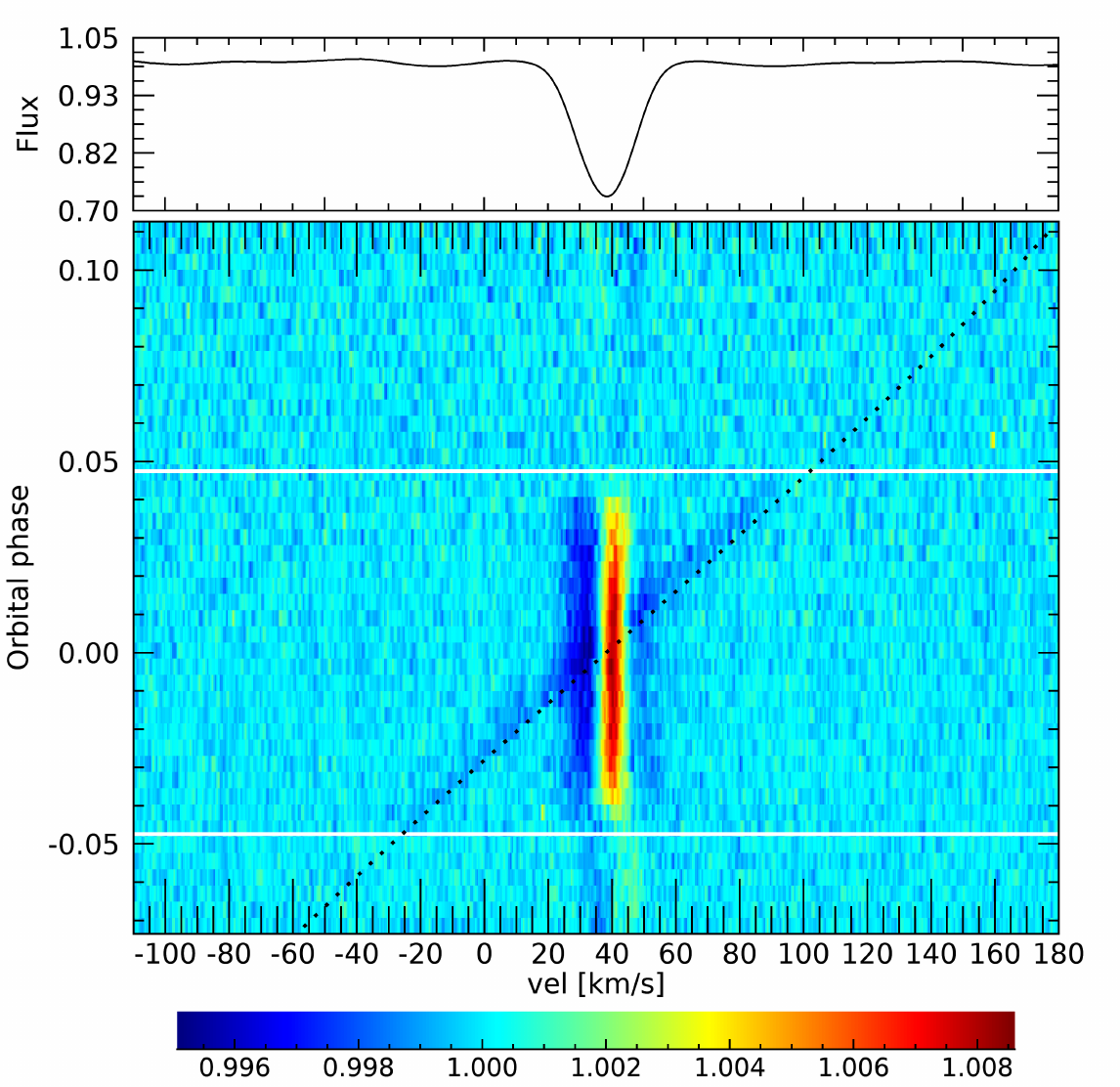}
\caption{{\it (Top panel)} Average out-of-transit stellar CCF. {\it (Bottom panel)} Tomography of CCF residuals for the 1-UT transit, in the stellar restframe. The stellar systemic velocity V$_{\rm sys}$ is still not subtracted. Horizontal white lines represent the beginning and end of the transit. The dotted black line shows the theoretical planetary RV. Differences between the stellar CCF residuals before and after the transit, possibly due to stellar activity, are noticeable.}
\label{fig.tomoCCF}
\end{figure}

The RVs were extracted with a F9 stellar mask using the DRS version 2.0.0, with a step of 0.5 \kms\ (1.0 \kms\ for the 4-UT data) and using a broad velocity range for the CCF [-150,150] \kms\ due both to the moderate projected rotational velocity of the star ($\sim$12 \kms, Sect. \ref{sec:stellar_par}) and to check for the possible presence of any planetary signal whose lines should fall several dozens of \kms\ away from the stellar line.
A list of the RVs is presented in Table \ref{tab:rv}.
At first look the RVs present a clearly anomalous in-transit curve (Fig. \ref{fig.rv}), which is not predicted by any model of classic RM effect \citep[e.g.,][]{2008ApJ...682.1283W}.
Looking together at the RV curve and the CCF residuals tomographic map (CCFs divided by an average out-of-transit CCF, Fig. \ref{fig.tomoCCF}), we investigated the possibility of an extreme case of atmospheric RM effect.

\renewcommand{\arraystretch}{0.8}
\begin{table}[!h]
\begin{center}
\caption{ESPRESSO RV observations of WASP-121.}
\label{tab:rv}
\footnotesize
\begin{tabular}{cccc}
 \hline\hline
 \noalign{\smallskip}
Time [BJD-2450000] & RV [\kms] & RV error [\ms] & Mode\\
 \noalign{\smallskip}
 \hline
\noalign{\smallskip}
8453.73466   & 38.1811   &     3.2& 4-UT\\
8453.73889   & 38.1719    &    3.1& 4-UT\\
8453.74314   & 38.1575     &   2.9& 4-UT\\
8453.74709   & 38.1541      &  3.1& 4-UT\\
8453.75150   & 38.1536     &   3.0& 4-UT\\
8453.75556   & 38.1579     &   2.8& 4-UT\\
8453.75972   & 38.1660     &   2.7& 4-UT\\
8453.76391   & 38.1783     &   2.6& 4-UT\\
8453.76804   & 38.1783     &   2.8& 4-UT\\
8453.77230  &  38.1649     &   2.8& 4-UT\\
8453.77636  &  38.1494     &  2.6& 4-UT\\
8453.78054  &  38.1326     &   2.6& 4-UT\\
8453.78466  &  38.1228     &   2.4& 4-UT\\
8453.78883  &  38.1219     &   2.5& 4-UT\\
8453.79297  &  38.1194     &   2.5& 4-UT\\
8453.79720  &  38.1308     &   2.7& 4-UT\\
8453.80133  &  38.1442     &   2.5& 4-UT\\
8453.80543  &  38.1658     &   2.6& 4-UT\\
8453.80959  &  38.1674     &   2.5& 4-UT\\
8453.81379  &  38.1671     &   2.7& 4-UT\\
8453.81790  &  38.1641     &   2.9& 4-UT\\
8453.82214  &  38.1597     &   2.9& 4-UT\\
8453.82636  &  38.1513     &   2.8& 4-UT\\
8453.83054  &  38.1542     &   2.8& 4-UT\\
8453.83462  &  38.1502     &   2.9& 4-UT\\
8453.83877  &  38.1450     &   2.9& 4-UT\\
8453.84302  &  38.1452     &   2.8& 4-UT\\
8453.84706  &  38.1367     &   2.6& 4-UT\\
8453.85128  &  38.1360     &  2.7& 4-UT\\
8490.63286  &  38.2674     &   5.6& 1-UT\\
8490.63824  &  38.2619     &   5.5& 1-UT\\
8490.64356  &  38.2535     &   5.5& 1-UT\\
8490.64904  &  38.2508     &   5.4& 1-UT\\
8490.65438  &  38.2446     &   5.5& 1-UT\\
8490.65977  &  38.2478     &   5.1& 1-UT\\
8490.66527  &  38.2391     &   5.7& 1-UT\\
8490.67063  &  38.2254     &   5.0& 1-UT\\
8490.67609  &  38.2147     &   5.3& 1-UT\\
8490.68146  &  38.2027     &   5.1& 1-UT\\
8490.68666  &  38.1982     &   5.8& 1-UT\\
8490.69221  &  38.1854     &  4.7& 1-UT\\
8490.69763  &  38.1806     &   5.0& 1-UT\\
8490.70312  &  38.1617     &   4.9& 1-UT\\
8490.70834  &  38.1460     &   5.2& 1-UT\\
8490.71383  &  38.1351     &   4.9& 1-UT\\
8490.71914  &  38.1328     &   5.3& 1-UT\\
8490.72452  &  38.1236     &   5.2& 1-UT\\
8490.72979  &  38.1304     &   4.8& 1-UT\\
8490.73552  &  38.1526     &   6.7& 1-UT\\
8490.74084  &  38.1511     &   5.4& 1-UT\\
8490.74618  &  38.1323     &   5.3& 1-UT\\
8490.75148  &  38.1210     &   6.2& 1-UT\\
8490.75699  &  38.1012     &   5.7& 1-UT\\
8490.76278  &  38.1094     &   6.9& 1-UT\\
8490.76778  &  38.1085     &   6.6& 1-UT\\
8490.77311  &  38.1203     &   6.0& 1-UT\\
8490.77848  &  38.1458     &   6.5& 1-UT\\
8490.78417  &  38.1484     &   6.6& 1-UT\\
8490.78917  &  38.1467     &   6.6& 1-UT\\
8490.79492  &  38.1474     &   6.5& 1-UT\\
8490.80032  &  38.1395     &   6.7& 1-UT\\
8490.80551  &  38.1404     &   7.2& 1-UT\\
8490.81100  &  38.1284     &   6.3& 1-UT\\
8490.81630  &  38.1295     &   6.9& 1-UT\\
8490.82171  &  38.1306     &   7.1& 1-UT\\
8490.82723  &  38.1267     &   7.0& 1-UT\\
8490.83255  &  38.1176     &   6.7& 1-UT\\
8490.83795  &  38.1199     &   7.2& 1-UT\\
8490.84344  &  38.1214     &   7.2& 1-UT\\
8490.84880  &  38.1070     &   7.4& 1-UT\\
8490.85412  &  38.1093     &   7.2& 1-UT\\
8490.85964  &  38.1033     &   7.9& 1-UT\\
\noalign{\smallskip}
 \hline
\end{tabular}
\end{center}
\end{table}

\subsection{Atmospheric Rossiter-McLaughlin effect\label{sec.atmoRM}}

The atmospheric RM effect \citep{atmoRM} is a deviation of the in-transit RVs from the classical RM effect that happens when the atmosphere of the planet is intercepted by the mask used to create the stellar CCFs. By studying the shape of this deviation we can measure the extension of the planetary atmosphere that correlates with the stellar mask. 
When comparing the 1-UT and 4-UT RVs, we find a difference in the RV slope (Fig. \ref{fig.rv}, bottom panel). Even including the RV offset given by the two different instrumental setups, this slope difference amounts to $\sim$20 \ms \ in $\sim$3 hours and has a 4.5$\sigma$ significance. When subtracting this slope (i.e., the linear trend fitted in Fig. \ref{fig.rv}, bottom panel), the two RV curves are fully compatible within the error bars. We cannot exclude that this is caused by the presence of additional bodies in the system, although \citet{bourrier} analyzed three transits with HARPS without finding evidence of any changing slope. One more possibility is the different levels of stellar activity of the star during the two transits (see Sect. \ref{sec:activity}). Since our aim is to  analyze the in-transit RVs, we detrend the data from the Keplerian motion by fitting a linear model on the out-of-transit RVs, and subtract the fit result from the data. We limit our atmospheric RM analysis of the RVs to the complete 1-UT transit. 
The incompleteness of the 4-UT transit is a major obstacle in constraining the whole set of parameters of our model. 

\begin{figure}
\centering
\includegraphics[width=\linewidth]{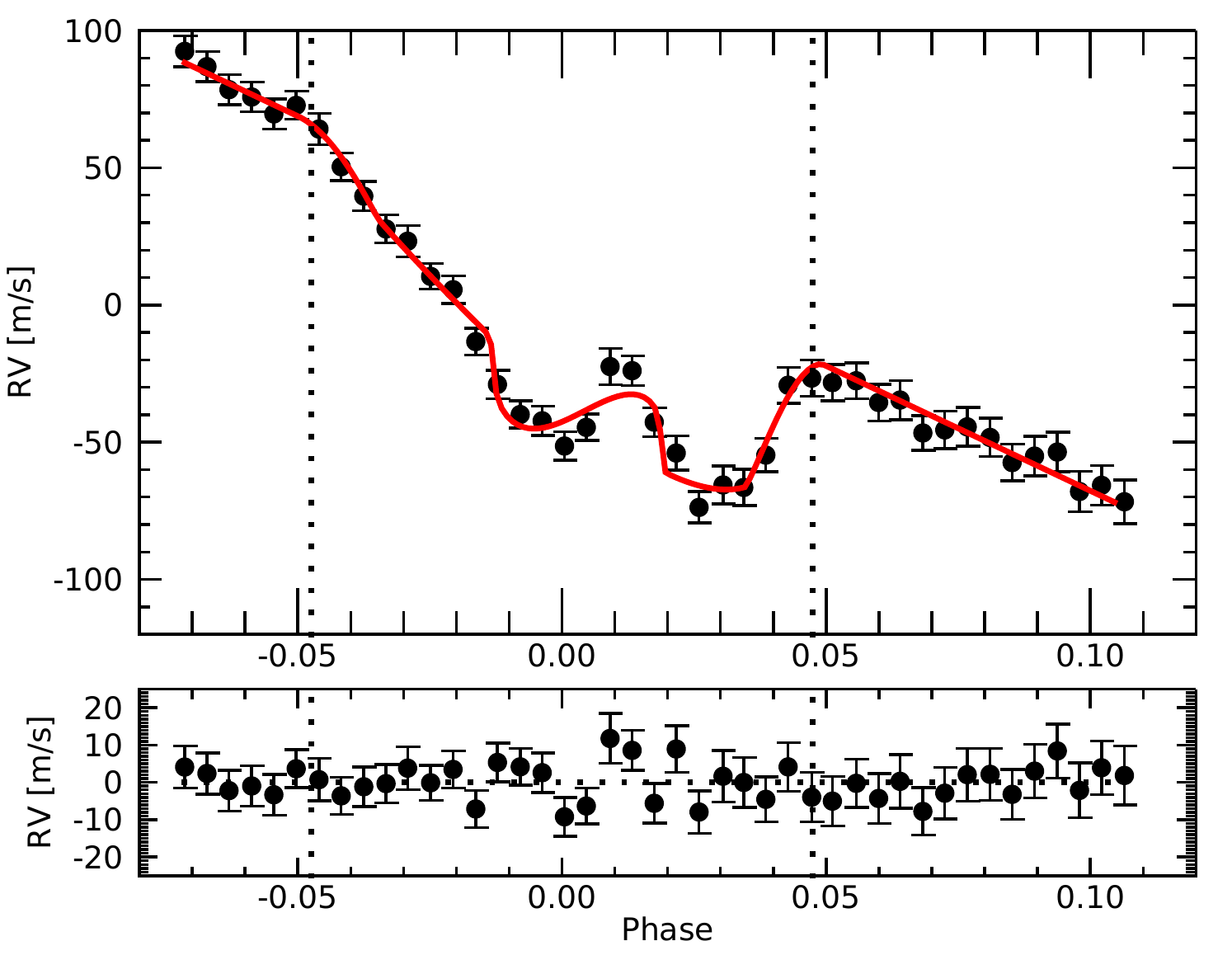}
\caption{RVs of the 1-UT transit together with the RM + atmospheric RM fit performed. The  residuals are shown in the bottom panel.}
\label{fig.RMfit}
\end{figure}

\begin{figure}
\centering
\includegraphics[width=\linewidth]{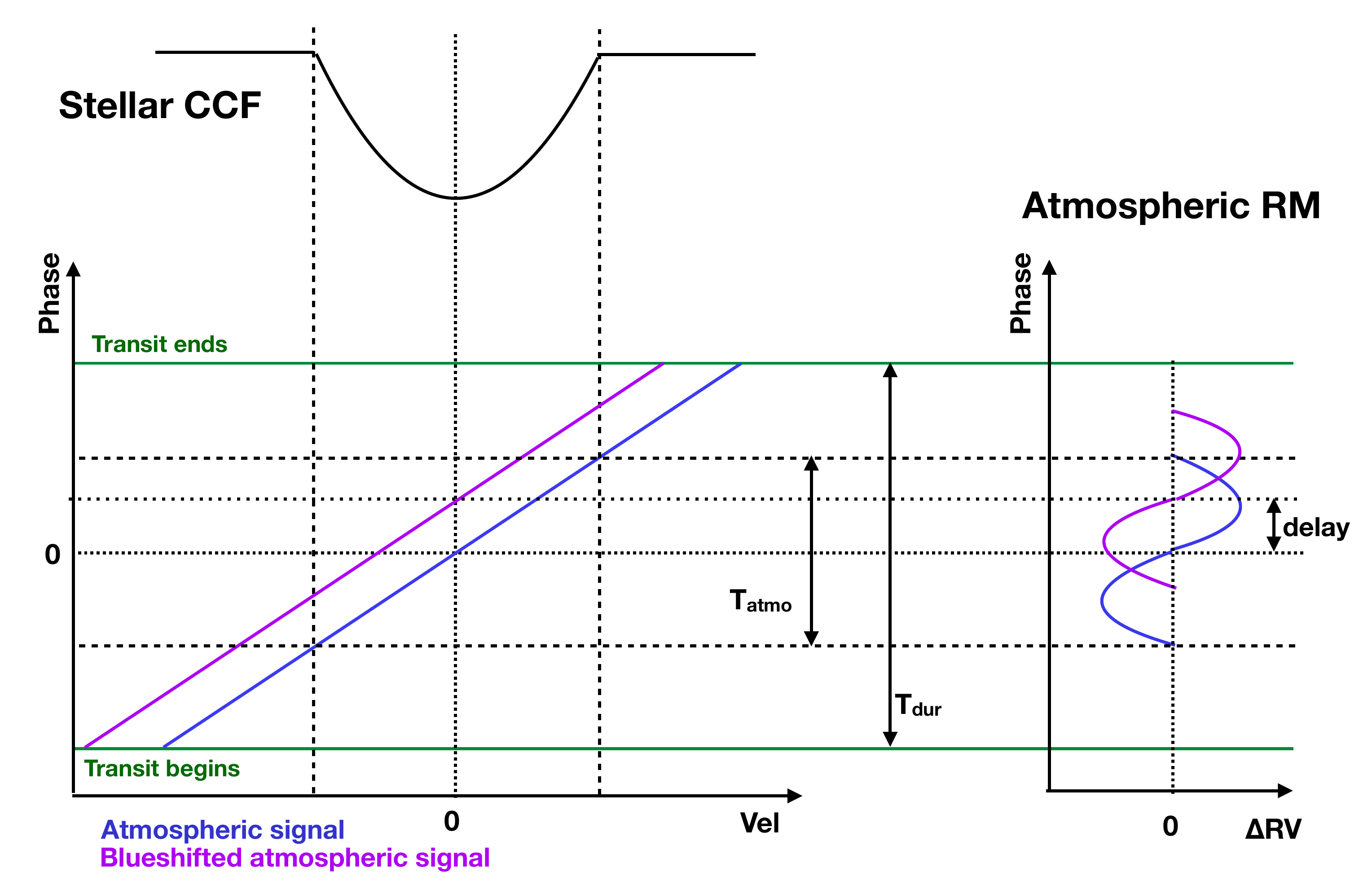}
\caption{ Schematic view of the atmospheric RM effect.}
\label{fig.schema}
\end{figure}

In \citet{atmoRM}, we first subtracted  the best fit RM model with literature parameters from the data, and then fitted the atmospheric RM.
Here instead we fit a model that is the sum of two RMs.
The first   is the classic RM \citep[taken from][]{ohta}, where as planet-to-star radius ratio (R$_{\rm p}/$R$_{\rm s}$) we use the sum of the R$_{\rm p}/$R$_{\rm s}$ calculated in the literature from the photometric transits and the R$_{\rm p,atmo}/$R$_{\rm s}$ described below. This is to take into account  the effects of the atmosphere on the classic RM.
An increase in the apparent R$_{\rm p}/$R$_{\rm s}$ because of the atmosphere is the same effect that causes the chromatic RM \citep{2004MNRAS.353L...1S,2009A&A...499..615D,2015A&A...580A..84D}, which in this case  affects the whole wavelength range and thus is not visible as wavelength dependent.
The second  is the atmospheric RM.
The atmospheric RM is basically a RM-like model, where the ratio R$_{\rm p,atmo}/$R$_{\rm s}$ is related to the extension of the atmosphere of the planet as if it were a disk. By definition, during a transit the planet (and thus the planetary atmosphere) travels from negative to positive RVs, following the Keplerian motion. If we consider that in the central part of the transit this can be  approximated well by a straight line, this is the same as the Doppler shadow of an aligned planet, but since the perturbation on the CCF of the atmosphere is opposite (i.e., an excess absorption, as opposed to the Doppler shadow)   its effect on the RVs will also be the opposite. We thus use a RM-like model (but with a different transit duration; see the description of the {\it ratio} parameter below) with the spin-orbit angle of the atmospheric track ($\lambda_{atmo}$) fixed to 180$^\circ$. 
We also include  other two parameters. The first is {\it delay}, which is related to the blueshift of the atmospheric signal. If the atmospheric track presents a blueshift, as  is often observed for exoplanets with transmission spectroscopy \citep[e.g.,][]{espresso1, bourrier}, the RV of the atmosphere at the center of the transit will not be at zero. This will cause a delay in the atmospheric RM, making its center to be slightly postponed with respect to the center of the transit. The {\it delay} parameter is thus the phase shift related to the average blueshift of the atmospheric signal.
The second parameter we include here is {\it ratio}, which is the ratio of the true a/R$_{\rm s}$ parameter (with $a$ the semi-major axis)  of the planet to the a/R$_{\rm s}$ that describes the passage of the atmospheric track over the stellar CCF. This is basically the ratio of the duration of the classic RM to that of the atmospheric RM (T$_{\rm dur}$/T$_{\rm atmo}$ in Fig. \ref{fig.schema}) as the transit duration is proportional to a/R$_{\rm s}$ \citep[e.g.,][]{2003ApJ...585.1038S}. We leave $ratio$ as a free parameter in the fit because we do not know a priori the width of the atmospheric signal, and so where its impact starts and ends at the edges of the CCF when extracting the RVs. We let the data tell us this in this way.
We note that in our model we are assuming for simplicity (and because of the limited duration of the transit) that the atmospheric signal is constant both in amplitude and blueshift during the transit.
A schematic view of the atmospheric RM and of the described parameters is presented in Fig. \ref{fig.schema}.

We fitted the RVs in a Bayesian framework by employing a differential evolution Markov chain Monte Carlo (DE-MCMC) technique \citep{TerBraak2006, Eastmanetal2013}, running five DE-MCMC chains of 100,000 steps and discarding the burn-in. 
The medians and the 15.86\% and 84.14\% quantiles of the posterior distributions were taken as the best
values and $1\sigma$ uncertainties. 
We fixed R$_{\rm p}/$R$_{\rm s}$, $i$, t$_0$, and a/R$_{\rm s}$ to the values listed in  Table~\ref{tabParameters}. We left as free parameters the linear limb darkening coefficient $\mu$, the spin-orbit angle $\lambda$, \vsini, the atmospheric R$_{\rm p,atmo}/$R$_{\rm s}$, $delay$, and $ratio$.
Priors for $\mu$ and $\lambda$ were set to the values in \citet{bourrier}, for \vsini\ to the value determined in Sect. \ref{sec:stellar_par}, and for the other three parameters to the result of a fit performed with the \texttt{MPFIT} IDL routine.

\begin{figure*}
\centering
\includegraphics[width=\linewidth]{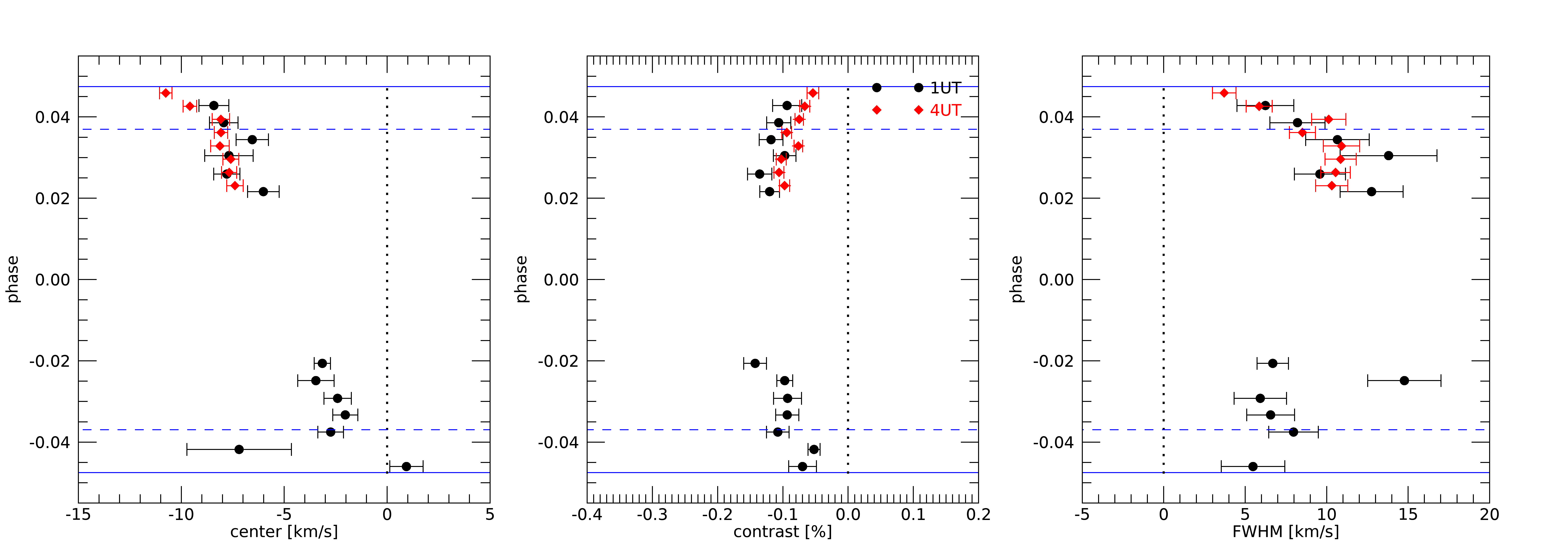}
\caption{Center {\it (left panel)}, contrast {\it (central panel),} and width {\it (right panel)} of the in-transit atmospheric CCFs Gaussian fits as a function of the orbital phase. Horizontal blue lines show the transit duration, dashed blue lines show the full-transit limits. Measurements are not done where the atmospheric track is superimposed on the Doppler shadow.}
\label{fig.atmotrack}
\end{figure*}

In Fig. \ref{fig.RMfit} we show the fit results and the residuals. We note that the two RV points that  deviate most  from the fit are those taken while the planetary atmospheric track overlaps the Doppler shadow, as is also seen  in the case of KELT-9b \citep{atmoRM}.
We find the sky-projected obliquity $\lambda= -87.08_{-0.27}^{+0.29}$, in agreement with \citet[][$\lambda= 87.2\pm 0.4$,  $-87.2\pm 0.4$ translated in our reference system]{bourrier}. 
The value of stellar \vsini=15.4$\pm$0.8 \kms\ is larger than that retrieved from spectral analysis, we note however that different RM models are known to produce \vsini\ measurements often in disagreement with each other and with estimates obtained from spectral line broadening \citep{brown}.
We can translate the $delay$=$0.0029_{-0.0008}^{+0.0005}$ into an average blueshift of the planetary atmospheric track $-4\pm1$ \kms. 
The value $ratio$=$2.61_{-0.13}^{+0.21}$ measures the ratio between the durations of the RM and atmospheric RM effects.
The atmospheric extension measurement is R$_{\rm p,atmo}/$R$_{\rm s}=-0.044_{-0.004}^{+0.003}$. 
Following \citet{atmoRM} from this value we measured the height of the planetary atmosphere correlating with the stellar mask, which is R$_{\rm atmo}$=1.052$\pm$0.015 R$_{\rm p}$.
This height represents  the extension of \ion{Fe}{i} in the atmosphere, as the stellar mask used to compute the CCF is mainly composed of \ion{Fe}{i} lines \citep[which represent $\sim$76\% of the weighted spectroscopic information of the mask;][]{espresso1}.
This value is in agreement with the results from \citet[][\ion{Fe}{i} present in low layers of the atmosphere]{gibson} and \citet[][R$_{\rm FeI}$ $\sim$1.03 R$_{\rm p}$]{cabot}, which measured \ion{Fe}{i} in the atmosphere by means of cross-correlation with theoretical templates.

Although the Doppler shadow is indeed larger than the atmospheric signal, we note that the RV deviation given by the atmospheric RM is of the same order of magnitude as the value given by the classic RM.
This can happen because the Doppler shadow in this orbital configuration (almost polar projected orbit) is always close to the center of the CCF, while the atmospheric signal moves along it from the far left to the far right, and (even if much smaller) this results in a RV deviation almost as large as the one given by the Doppler shadow.

\subsection{Planetary CCF\label{sec:planetaryCCF}}

The planetary atmospheric track shown in Fig. \ref{fig.tomoCCF} is slightly shifted from the theoretical planetary velocity computed with the orbital parameters of Table \ref{tabParameters}. ESPRESSO has recently proven its capability to resolve time variations of this atmospheric track \citep{espresso1}.
For WASP-121b, \citet{bourrier} found a similar   behavior. We thus investigated this with our data, by fitting a Gaussian profile to each atmospheric line profile in the planetary restframe. Errors on the CCF residuals were set to the standard deviation of their continuum. We excluded the part of the transit where the atmospheric track overlaps the Doppler shadow \citep[see][]{espresso1}. The results are shown in Fig. \ref{fig.atmotrack}. The velocity center of the atmospheric track changes with time (Fig. \ref{fig.atmotrack}, left panel), becoming more blueshifted in the second part of the transit, in a way similar to what happens in the case of the ultra-hot gas giant WASP-76b \citep{espresso1}. This is confirmed by the agreement between the 1-UT and 4-UT data, unfortunately only possible for the second part of the transit.
We measure an average change in blueshift from $-2.80\pm$0.28\kms\ in the first half to $-7.66\pm$0.16\kms\ in the second half, excluding the ingress and egress. 
 The results are close to those of \citet{bourrier}, who discovered that the atmospheric signal was becoming more blueshifted from the first to the second part of the transit.
 While we find no differences in the atmospheric CCF contrast ($1050\pm$80 versus $970\pm$30 ppm), its width increases but not significantly ($7.3\pm$0.7 versus $10.2\pm$0.4\kms).
The contrasts can be translated into an effective planetary radius of $\sim$1.03 R$_{\rm p}$ (see details in Sect. \ref{sec:discussion}), which is in agreement with the R$_{\rm atmo}$ found with the RV analysis (R$_{\rm atmo}$=1.052$\pm$0.015 R$_{\rm p}$), thus reinforcing the validity of the atmospheric RM method. The average blueshift values found with the two analyses are also consistent with each other.


\section{Transmission spectrum\label{sec:trans_spectrum} }

\begin{figure*}[!h]
\centering
\includegraphics[width=\linewidth]{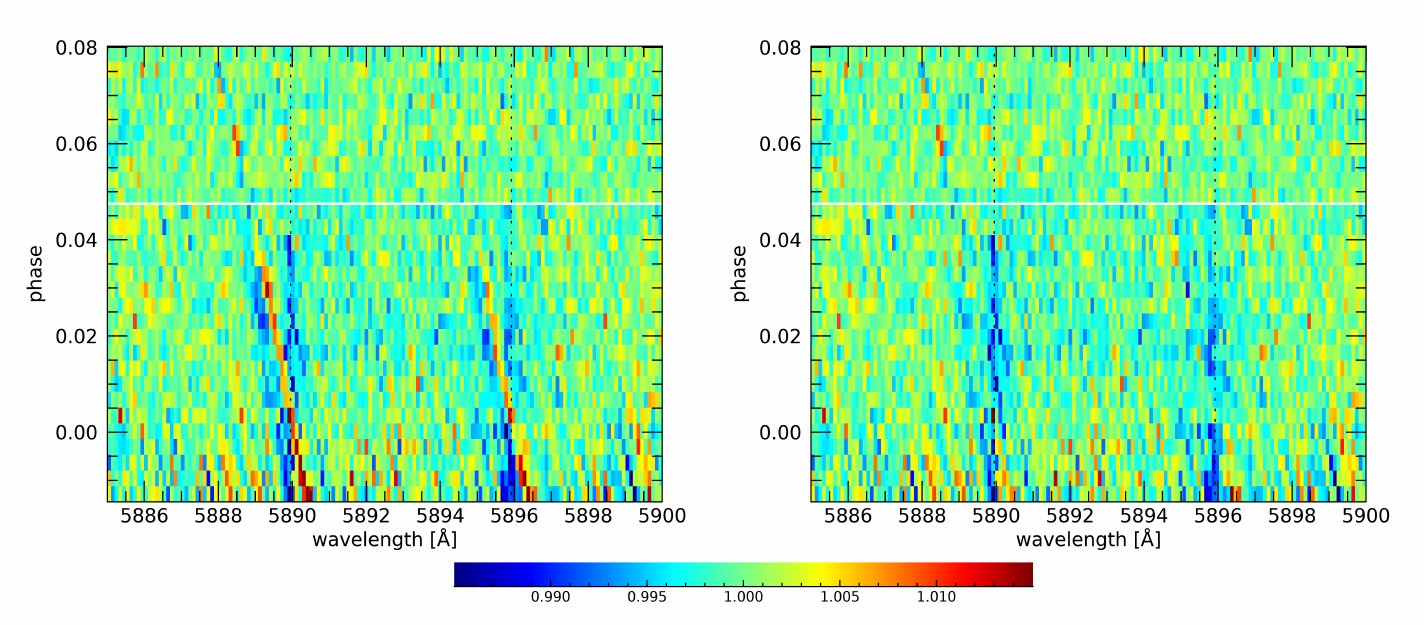}
\caption{Tomographic map of the sodium D doublet in the planetary restframe for the 4-UT transit, shown before (left panel) and after (right panel) the correction for the CLV and RM effects.
The horizontal white line shows the end of the transit; the vertical dashed lines represent the planetary restframe of the sodium D lines.}
\label{fig.clvtomo}
\end{figure*}

\begin{figure}[!h]
\centering
\includegraphics[width=\linewidth]{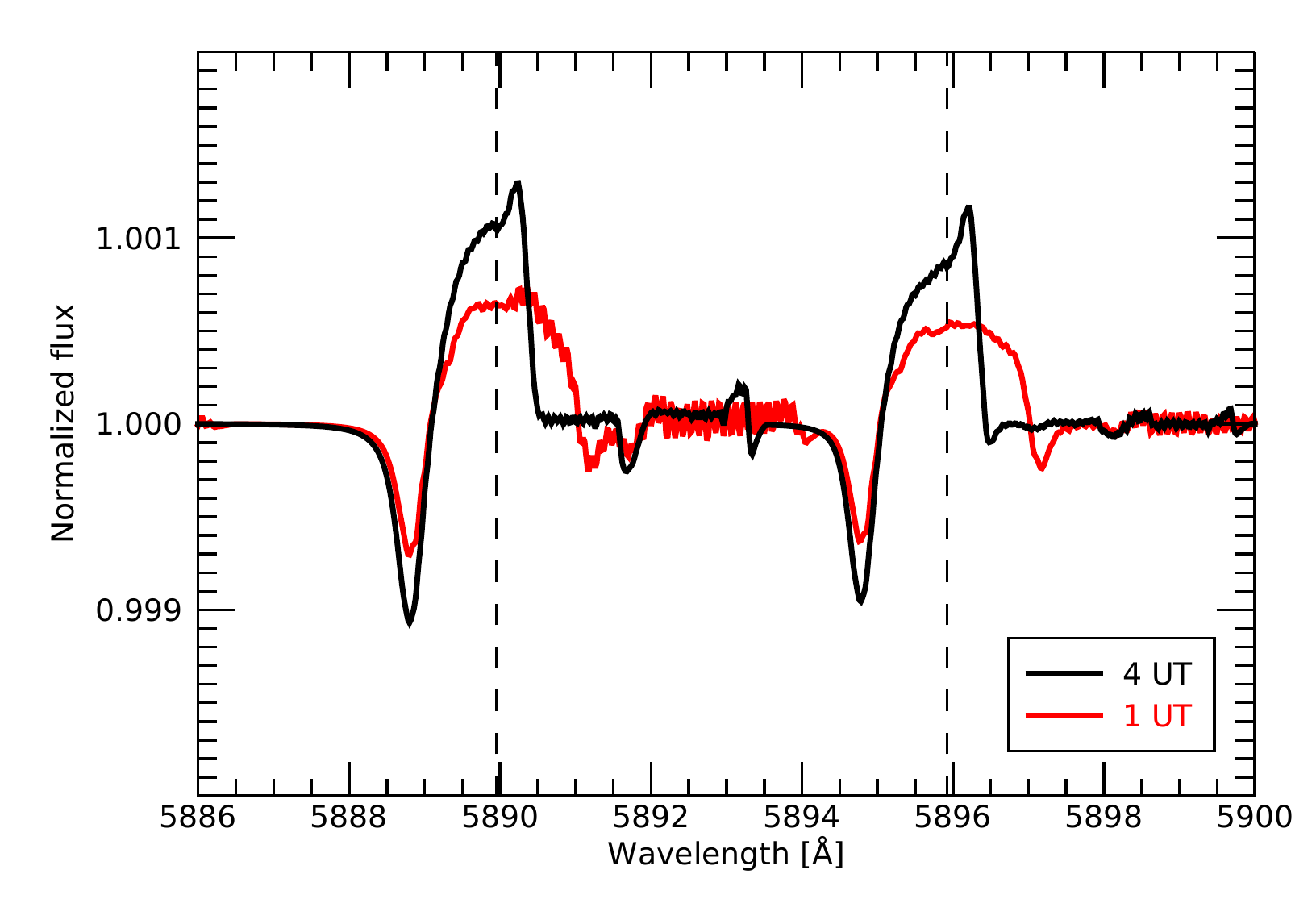}
\caption{Impact of CLV+RM effects on the transmission spectra of 1-UT (red line) and 4-UT (black line) observed transits in the wavelength region of the sodium D doublet.}
\label{fig.clv1vs4}
\end{figure}

The transmission spectrum is extracted following basically the procedure of \citet{wyttenbach}, which was performed independently for each of the two transits.
We first shift the spectra to the stellar restframe by using the Keplerian model of the system with the parameters of Table \ref{tabParameters}, then we normalized each spectrum.
Telluric correction is performed by exploiting the scaling relation between airmass and telluric line strength \citep{2008A&A...487..357S,2010A&A...523A..57V,2013A&A...557A..56A}, rescaling all the spectra as if they were observed to the airmass of the transit center. We can perform telluric correction in the stellar restframe as the variation of the Barycentric Earth RVs during one night is well below the resolution of the instrument. Then we divided all the spectra by a master stellar spectrum created by averaging all the out-of-transit spectra.
At this stage we find in the residual spectra a clear pattern of  {wiggles}, which was already described in \citet{tabernero_esp}. We correct for this pattern by fitting for a sinusoid with varying period, phase, and amplitude for each spectrum, with the fit performed independently for each wavelength region where we looked for spectral features.
 As a final step we move to the planetary restframe by shifting all the residual spectra for the theoretical planetary radial velocity (calculated with values from Table \ref{tabParameters}). The transmission spectrum is now created by averaging all the full-in-transit residual spectra.
 We note again that the spectral resolution is different for the two transits analyzed. In order to take this into account when estimating the significance of the detections in the transmission spectrum, the final rebinning is done with a wavelength step of 0.01 $\AA$ for the 1-UT data and of 0.02 $\AA$ for the 4-UT data. These values were chosen because they are very close to the mean resolution step calculated on the whole spectrograph range in the two observing modes.

\subsection{Stellar contamination in the transmission spectrum\label{sec:clv+rm}}
The star in front of which the planet transits is not a simple homogeneous disk, but rotates and has a surface brightness that changes as a function of the distance from the center.
Effects such as center-to-limb variation (CLV) and stellar rotation bring spurious signals in the transmission spectrum, possibly causing false detections \citep[e.g.,][]{2020arXiv200210595C} and incorrect line-profile estimations \citep[e.g.,][]{2018A&A...617A.134B}.

For the case of the WASP-121b transmission spectrum, these effects have been proven to be negligible while analyzing HARPS data \citep{cabot,bourrier}. Since with ESPRESSO we clearly see their impact in the 4-UT spectra (Fig. \ref{fig.clvtomo}), we take them into account when analyzing our data. We thus created a model following the methodology of \citet{yan}. 
The star is modeled as a disk divided in sections of 0.01 \rs. For each point we calculate the $\mu$ value ($\mu=\cos{\theta}$, with $\theta$ the angle between the normal to the stellar surface and the considered line of sight) and the projected rotational velocity (by rescaling the \vsini\ value of Table \ref{tabParameters}).
Then we assign a spectrum to each point of the grid, by quadratically interpolating on $\mu$ and Doppler-shifting the model spectra created using the tool Spectroscopy Made Easy \citep[SME,][]{2017A&A...597A..16P}, with ATLAS9 
stellar atmospheric models (assuming solar abundances and local thermodynamic equilibrium approximation) and the line list from the VALD database \citep{2015PhyS...90e4005R}. The model spectra are created for 21 different $\mu$ values and with null rotational velocity, and adapted to the resolving power of the instrument.
Then we simulate the transit of the planet, calculating for different orbital phases the stellar spectrum as the average spectrum of the non-occulted modeled sections.
We divide these for an average out-of-transit stellar spectrum, and we have as a result the simulated RM+CLV effects at each in-transit orbital phase.
In Fig. \ref{fig.clvtomo} (right panel) we show in a 2D tomographic map the correction with our model applied to the 4-UT data in the wavelength zone of the sodium D doublet.
We can calculate the impact on the average transmission spectrum by moving all the in-transit spectra in the planetary restframe and averaging them.

During the transit of a planet the occulted stellar regions are different. This means that the analysis of planetary transmission spectra which are calculated from complete or incomplete transits will be affected in a different way by the CLV and RM effects. In Fig. \ref{fig.clv1vs4} we show their impact on the sodium D doublet wavelength region of the two ESPRESSO transmission spectra analyzed. The system orbital configuration makes the global effect larger by a factor $\sim$2 for the 4-UT partial transit.

\begin{figure}[]
\centering
\includegraphics[width=\linewidth]{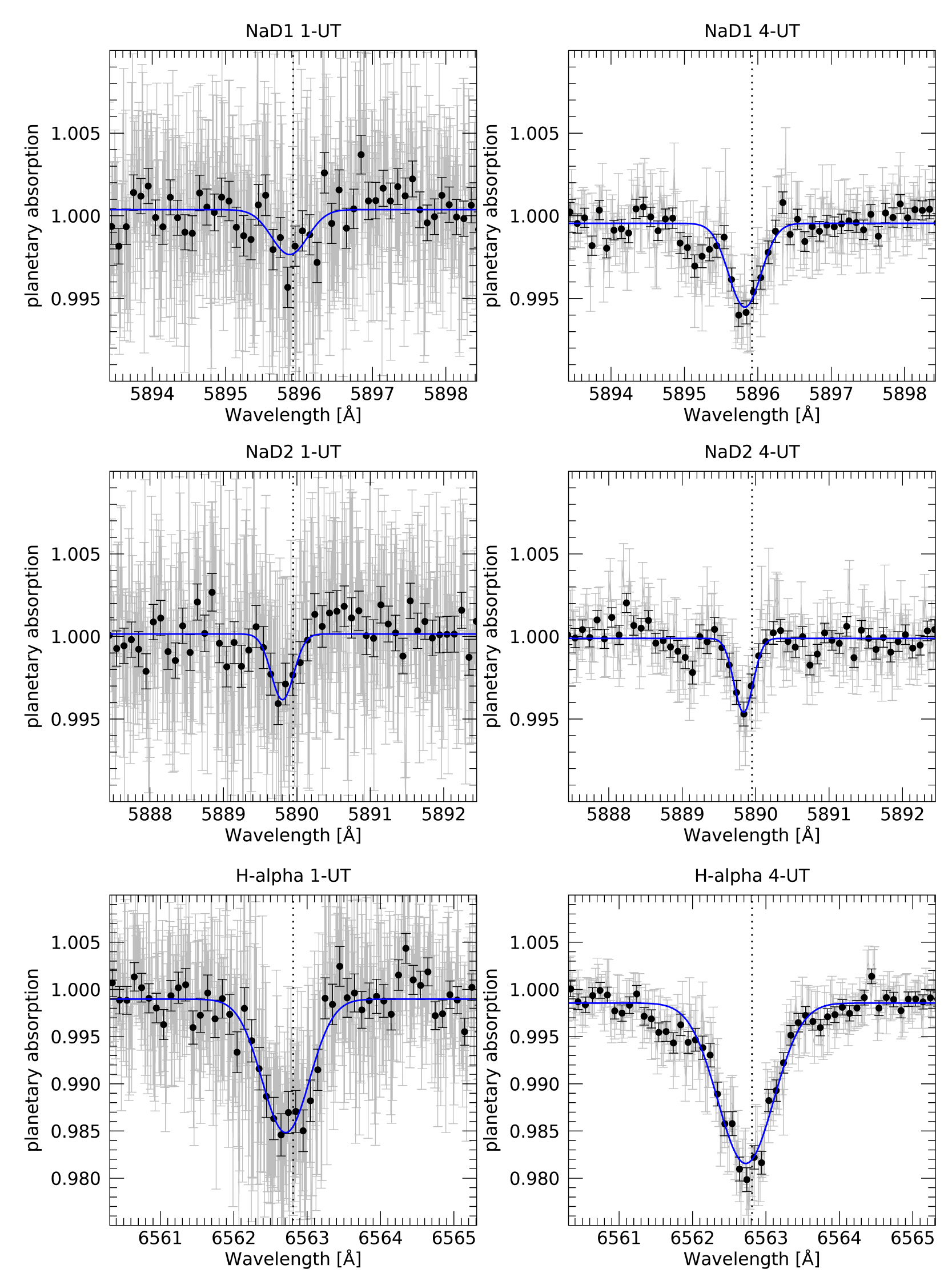}
\caption{Detections of Na D1 (first row), Na D2 (second row), and H$\alpha$ (third row) for the transits observed with 1-UT (first column) and 4-UT (second column).
The vertical scale is the same for each row, showing the precision of the 4-UT data.
The black points represent 0.1 $\AA$ binning; the blue line is the best fit Gaussian profile. The vertical dotted line shows the expected planetary restframe.}
\label{fig.detections_literature}
\end{figure}

\begin{figure}[]
\centering
\includegraphics[width=\linewidth]{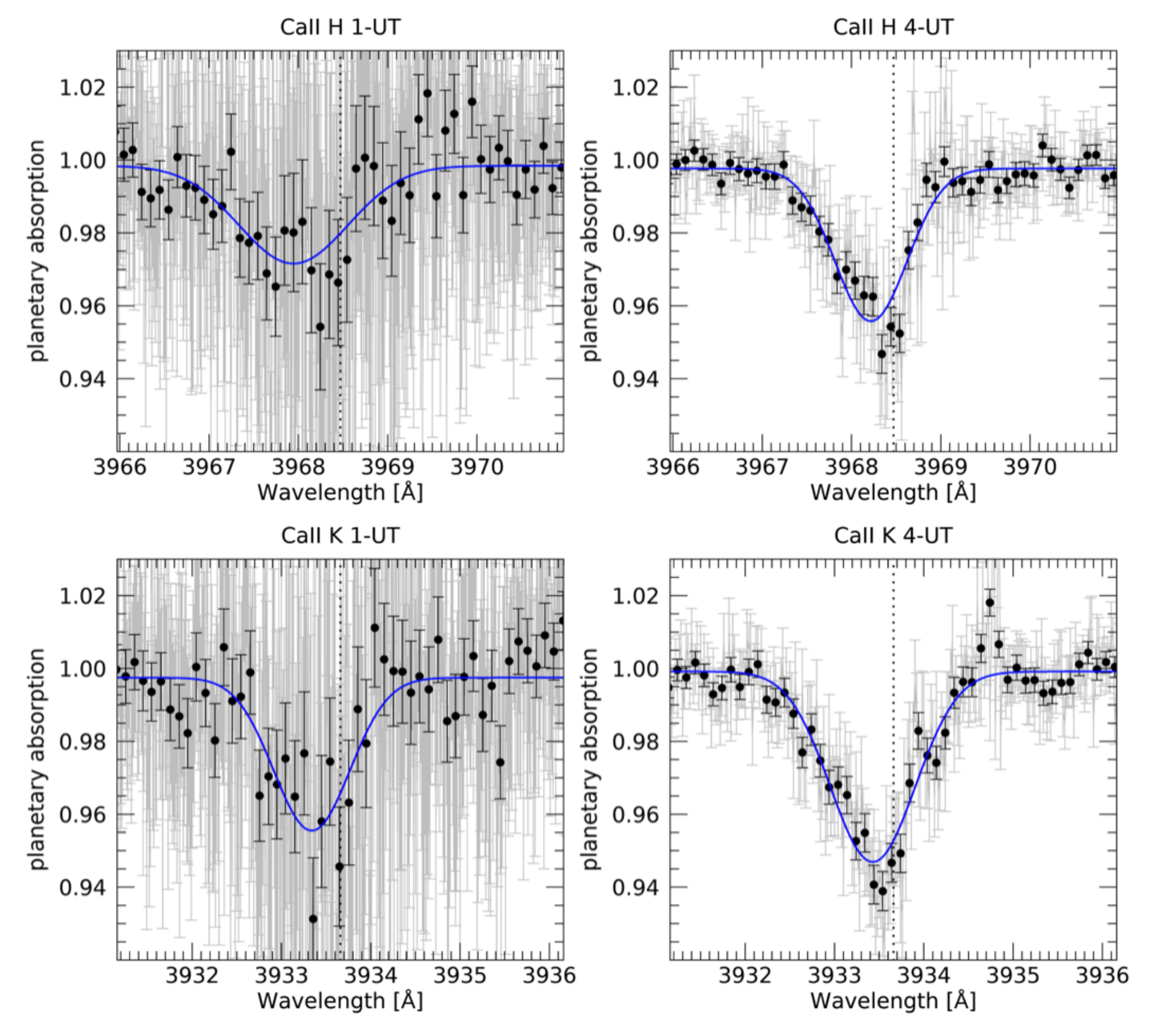}
\caption{New detections of \ion{Ca}{ii} H (first row) and \ion{Ca}{ii} K (second row) in the 1-UT and 4-UT transits (first and second column, respectively).
The black points represent 0.1 $\AA$ binning; the blue line is the best fit Gaussian profile. The vertical dotted line shows the expected planetary restframe.}
\label{fig.Ca_lines}
\end{figure}

We note that \citet{cabot} showed the non-impact of stellar effects CLV+RM on the transmission spectrum of WASP-121b retrieved using three HARPS transits. Our model is consistent with theirs, but given the higher quality of our ESPRESSO data (and in particular the high S/N of the 4-UT transit; see  Fig. \ref{fig.clvtomo}) we chose to remove stellar effects from the data.
For uniformity in the analysis, we subtracted the CLV+RM model from the transmission spectra of both transits.

\subsection{Detections of planetary absorption lines\label{sec:detections}}

\begin{figure}
\centering
\includegraphics[width=\linewidth]{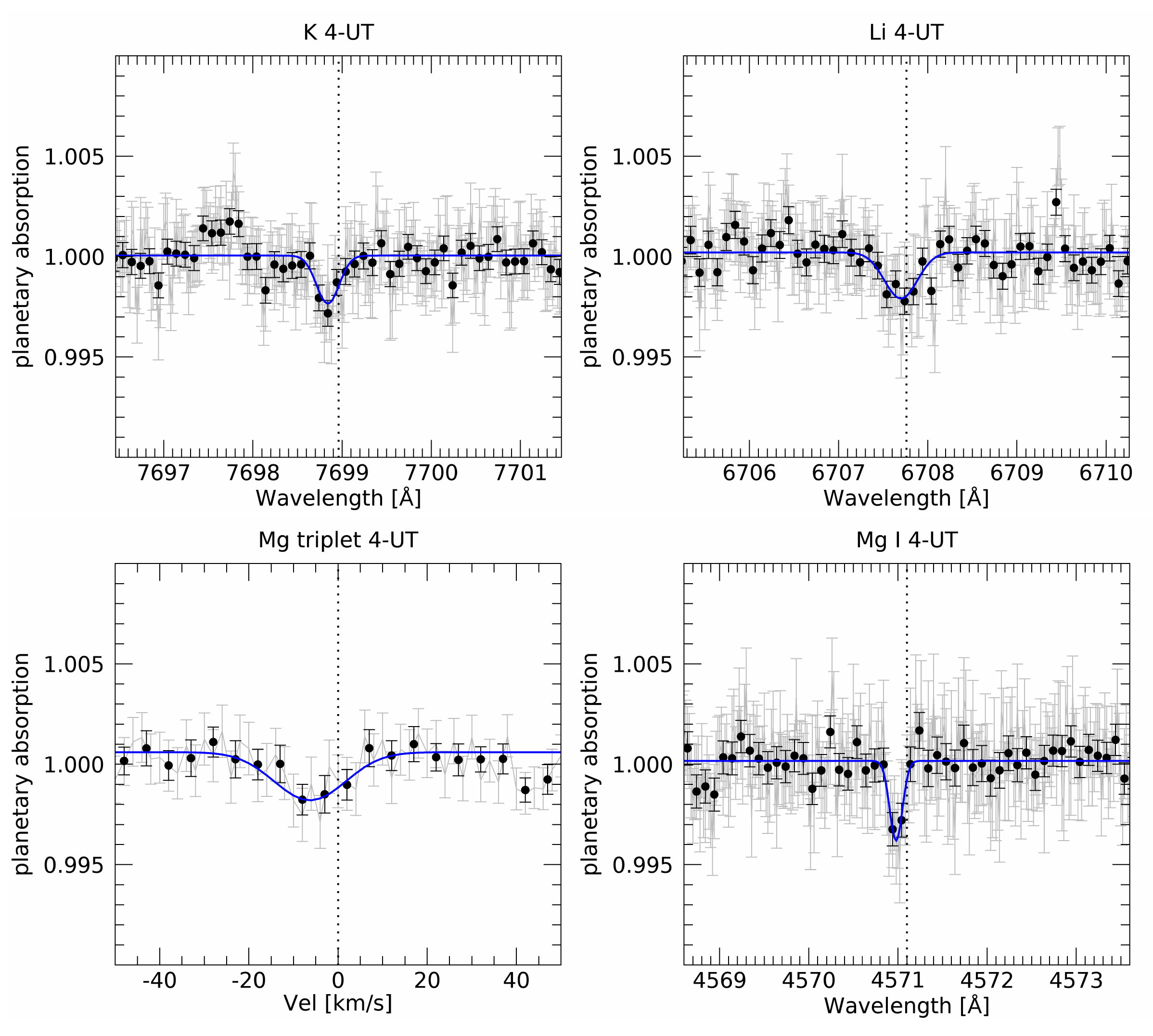}
\caption{New detections of K (top left), Li (top right), and Mg (bottom panels) in the 4-UT transit.
The black points represent 0.1 $\AA$ (5 \kms\ for Mg triplet) binning; the blue line is the best fit Gaussian profile. The vertical dotted line shows the expected planetary restframe.}
\label{fig.newdetections}
\end{figure}

\begin{figure}
\centering
\includegraphics[width=\linewidth]{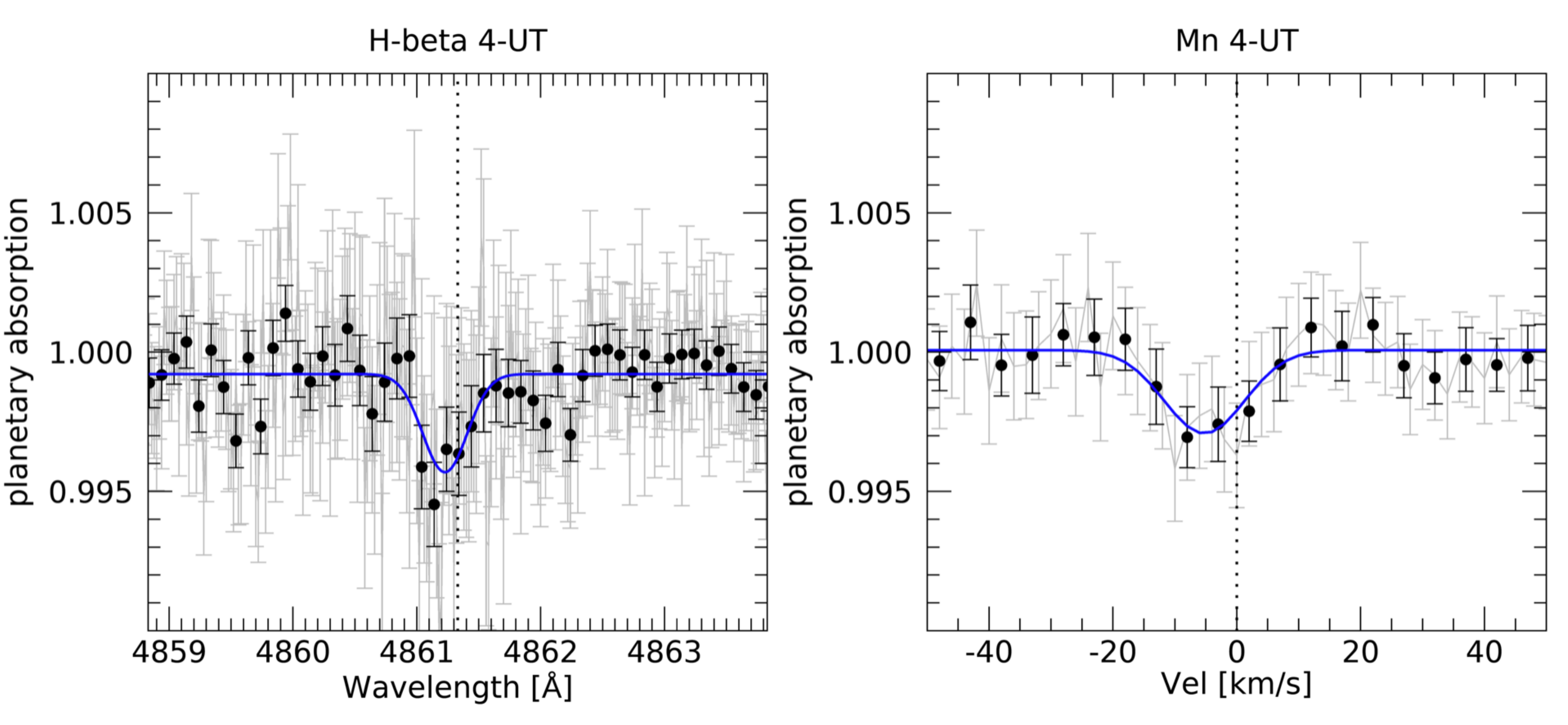}
\caption{Possible detections of H$\beta$ and Mn triplet (averaged in the velocity space) planetary line profiles in the 4-UT transmission spectrum.
The black points represent 0.1 $\AA$ binning (5 \kms\ for Mn); the  blue line is the best fit Gaussian profile. The vertical dotted line shows the expected planetary restframe.}
\label{fig.nonsignificant}
\end{figure}

We analyzed the transmission spectra retrieved in the two transits independently after the correction of the stellar RM+CLV effects because of the different resolving power of the two observing modes used and because of the difference in the stellar activity level (Sect. \ref{sec:activity}).
A summary of the absorption lines detected in the planetary atmosphere is reported in Table \ref{tabdetections}. The parameters are calculated with a Gaussian fit and a MCMC error bar estimation.
We report from the analysis of the two transits significant detections ($>4\sigma$)  for the previously detected Na D1, Na D2, and H$\alpha$ lines (Fig. \ref{fig.detections_literature}), together with new detections of \ion{Ca}{ii} H\&K (Fig. \ref{fig.Ca_lines}). Only in the 4-UT transit we can also detect  Li, \ion{Mg}{i}, and K (Fig. \ref{fig.newdetections}). We can detect only one line of the K doublet at $\sim$770 nm (Table \ref{tabdetections}), as the other one falls in a region of very strong telluric absorption.
The Mg triplet is detected only when averaging  the three lines in the velocity space.
We also report on possible detections of H$\beta$ (3.2$\sigma$) and Mn triplet (3.3$\sigma$) (Fig. \ref{fig.nonsignificant}).

\begin{table*}
\begin{center}
\caption{Summary of the detections in the transmission spectra of the two transits. 
 Detection is considered positive (Y) only at the level of 4$\sigma$ and beyond, and partial (YN) in the range 2.5-3.9$\sigma$. ``2D'' refers to the visibility in the 2D tomographic map.
}
\label{tabdetections}
\footnotesize
\setlength{\tabcolsep}{4pt}
\begin{tabular}{cc|cccc|cccc|c}
 \hline\hline
 \noalign{\smallskip}
 &  & \multicolumn{4}{c|}{1-UT} & \multicolumn{4}{c|}{4-UT} &\\
Element & $\lambda_{\rm air}$  & Contrast  & FWHM  & Shift  & Detection & Contrast  & FWHM & Shift  & Detection & 2D\\
 & [$\AA$] &  [\%] &  [\kms] &  [\kms] &  &  [\%] &  [\kms] &  [\kms]  &  & \\
 \noalign{\smallskip}
 \hline
\hline
\noalign{\smallskip}
\ion{Ca}{ii} K & 3933.66 & $4.20 \pm  0.70$ &$76.63 \pm 14.63$ & $-24.40 \pm   5.27$ &Y &$5.22 \pm   0.17$ &$85.1 \pm    3.1$&$-17.6 \pm  1.4$& Y& Y\\
\noalign{\smallskip}
\hline
\noalign{\smallskip}
\ion{Ca}{ii} H & 3968.47 & $2.68 \pm  0.51$ & $108.73 \pm 18.87$ & $-39.67 \pm 7.86$ &Y & $4.18 \pm 0.19$ & $71.9 \pm  3.7$ & $-19.0 \pm   1.6$ & Y&Y\\
\noalign{\smallskip}
\hline
\noalign{\smallskip}
& 4030.76 & & & & & & & & &\\
Mn & 4033.07&N/A &N/A&N/A & N&$0.29 \pm   0.08$ &$15.3 \pm  3.9$&$-5.2 \pm   2.1$& YN & N\\
& 4034.49 & & & & & & & & & \\
\noalign{\smallskip}
\hline
\noalign{\smallskip}
 Mg I & 4571.10   &N/A & N/A &N/A &N &$-0.40 \pm 0.08$ & $10.9 \pm 1.9 $& $-7.9 \pm   1.0$&Y&N\\
\noalign{\smallskip}
\hline
\noalign{\smallskip}
H$\beta$ &4861.33 &N/A & N/A &N/A &N &$0.35 \pm    0.11$ &$26.1 \pm     14.9$&$-6.3 \pm       4.3$&YN & YN\\
\noalign{\smallskip}
\hline
\noalign{\smallskip}
& 5167.32 & & & & & & & & &\\
Mg I & 5172.68&N/A &N/A& N/A& N&$-0.24 \pm   0.06$ &$13.4 \pm  4.4$&$-6.3 \pm  1.8$& Y & Y\\
& 5183.60 & & & & & & & & & \\
\noalign{\smallskip}
\hline
\noalign{\smallskip}
Na D2 & 5889.95 &$0.398 \pm   0.081$ &$18.42 \pm       4.56$ &$-7.42 \pm       1.98$&Y &$0.447 \pm   0.045$ &$15.6 \pm       1.9$& $-5.9 \pm      0.8$&Y & Y\\
\noalign{\smallskip}
\hline
\noalign{\smallskip}
Na D1 &5895.92 & $0.271 \pm   0.067$ & $29.53 \pm       9.94$ &$-2.45 \pm       3.18$&Y &$-0.51 \pm   0.048$ &$26.3 \pm       3.3$ &$-4.9 \pm    1.1$& Y& Y\\
\noalign{\smallskip}
\hline
\noalign{\smallskip}
H$\alpha$ &6562.81 &$1.414 \pm   0.090$ &$34.70 \pm       2.65$& $-4.64 \pm       1.18$&Y &$1.70 \pm   0.048$ &$40.9 \pm       1.7$&$-3.9 \pm      0.6$&Y & Y\\
\noalign{\smallskip}
\hline
\noalign{\smallskip}
Li & 6707.76&N/A &N/A& N/A & N &$0.23 \pm   0.04$ &$19.4 \pm       3.3$&$-2.7 \pm       1.6$ &Y &Y\\
\noalign{\smallskip}
\hline
\noalign{\smallskip}
K & 7698.96 &N/A &N/A& N/A& N &$0.23\pm0.05$ &$12.5\pm5.4$&$-4.9\pm1.2$&Y & N\\
\noalign{\smallskip}
\hline
 \hline
\end{tabular}
\end{center}
\end{table*}

\begin{figure*}
\centering
\includegraphics[width=\linewidth]{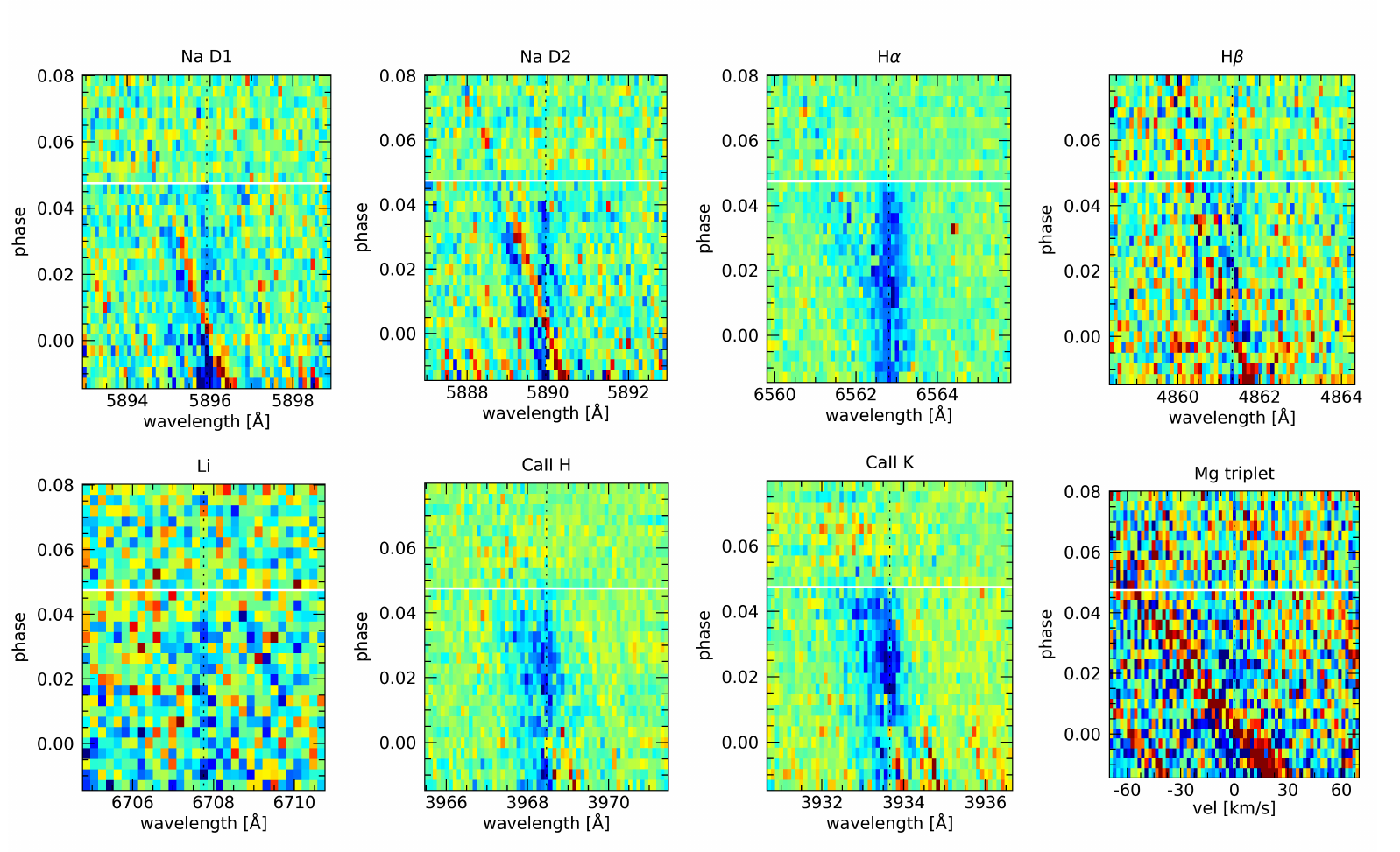}
\caption{Tomography positive quality checks of the detections in the planetary restframe, for the 4-UT partial transit, before applying the stellar contamination correction (this also shows  the restframe of the Doppler shadow, i.e., the red track). Each square represents a 0.1 \AA\ bin on the horizontal scale (except 0.2 \AA\ for Li and 1\kms\ for Mg). The color scale (not shown here as   these are qualitative plots) is different for each plot, from blue representing absorption to red representing emission. The vertical dotted lines represent the laboratory wavelength. The horizontal white lines represent the end of the transit. The Mg tomography (bottom right plot) is presented in the velocity space as the sum of the three lines of the triplet.}
\label{fig.tomo_all}
\end{figure*}

Checking the reference frame of the detections is mandatory when we want to be sure they are caused by the planetary atmosphere and not by spurious stellar effects \citep[e.g.,][]{2016ApJ...817..106B,2018A&A...617A.134B}. For each element we thus also report  whether the detection is resolved in the 2D tomographic map (Table \ref{tabdetections} and Fig. \ref{fig.tomo_all}). This quality check confirms that the absorption is in the planetary restframe for all the significant detections except for K, which is in a wavelength region particularly affected by strong telluric contamination and so its restframe cannot be discriminated in the 2D map.

Each element line detected in the transmission spectra presents a net blueshift, that for the low layers of the atmosphere (thus excluding \ion{Ca}{ii} H\&K and H$\alpha$ lines) on average is of $\sim$-5 \kms, which is indicative of winds in the planetary atmosphere coming from the dayside to the nightside.

\subsection{Cross-correlation with templates\label{sec:ccf_templates}}

We created WASP-121b high-resolution theoretical transmission spectra, expressed in planetary radius as a function of wavelength for different species of interest.
Models of \ion{Fe}{i}, \ion{Fe}{ii}, \ion{Ti}{i}, \ion{Ti}{ii}, \ion{Cr}{i}, \ion{V}{i}, TiO (Plez linelist), and VO (Plez linelist) were generated by using \texttt{petitRADTRANS} \citep{pRT}, while those of CaH and ZrO were generated via  \texttt{TURBOSPECTRUM} \citep{2012ascl.soft05004P}. We assumed solar abundances, an isothermal atmospheric profile with T=3000 K and a continuum pressure level of 1 mbar. These parameters were set because they are typical for UHJs \citep[e.g.,][]{2019arXiv190502096H,stangret}.
The atmospheric models were translated into flux (R$_{\rm p}$/R$_{\rm s}$)$^2$, convolved at the ESPRESSO resolving power and continuum normalized. 

Cross-correlation between the data and the models is performed in the stellar restframe on single residual spectra after the removal of a master out-of-transit star and telluric contamination (with the procedures explained in Sect. \ref{sec:trans_spectrum}) over the whole ESPRESSO wavelength range. 
We define the cross-correlation as 

\begin{equation}
C(v,t)= \sum\limits_{i=1}^N x_i(t,v) M_i
\label{equazccf}
,\end{equation}
where $M$ is the model normalized to unity and $x$ are the $N$ wavelengths of the spectra taken at the time $t$ and shifted at the velocity $v$.
In this way we preserve the flux information \citep[e.g.,][]{2019arXiv190502096H}. 
In our analysis we set  to zero all the model lines with contrast of less than 5\% of the maximum in our wavelength range.

\begin{table*}
\begin{center}
\caption{Summary of cross-correlation detections in the atmosphere of WASP-121b, with results for the 1-UT and 4-UT transits. Center and S/N are estimated for uniformity at the theoretical K$_{\rm p}$ value.}  
\label{tab:crosscorr}
\footnotesize
\begin{tabular}{c|ccc|ccc}
 \hline\hline
  \noalign{\smallskip}
 &   \multicolumn{3}{c|}{1-UT} & \multicolumn{3}{c}{4-UT} \\
 \noalign{\smallskip}
Element & K$_{\rm p}$ [km/s] & center [km/s] & S/N@K$_{\rm p, theoretical}$ &  K$_{\rm p}$ [km/s] & center [km/s] & S/N@K$_{\rm p, theoretical}$ \\ 
 \noalign{\smallskip}
 \hline
\noalign{\smallskip}
 \ion{Fe}{i} &  $203_{-9}^{+8}$  & $-7.5\pm0.3 $ & $13.3\pm0.6 $& $208_{-12}^{+18}$ &$-8.5\pm0.5$& $8.6\pm0.6$\\
  \noalign{\smallskip}
\ion{Fe}{ii}  & $195_{-12}^{+12}$   & $-6.0\pm0.8$& $4.4\pm0.4$ & N/A & N/A &N/A \\
 \noalign{\smallskip}
\ion{Cr}{i} &  $203_{-9}^{+8}$ & $-7.2\pm0.7$& $4.2\pm0.3$& $206_{-13}^{+11}$  &$-9.1\pm0.3$& $9.2\pm0.3$\\
 \noalign{\smallskip}
\ion{V}{i} &  $197_{-7}^{+11}$ & $-7.2\pm0.4$& $6.6\pm0.3$& $214_{-33}^{+23}$  &$-8.3\pm0.5$& $5.5\pm0.3$\\
\noalign{\smallskip}
 \hline
\noalign{\smallskip}
\end{tabular}
\end{center}
\end{table*}

We selected a step of 0.5 \kms\ (1 \kms\ for the 4-UT) and a velocity range [-200,200] \kms. 
The spectra are divided into segments of 200 $\AA$ \citep[see, e.g.,][]{2019arXiv190502096H}, then the cross-correlation is performed for each segment. 
We performed a sigma-clipping at 5$\sigma$ and masked the wavelength ranges most affected by telluric contamination (the intervals 5240-5280 \AA, 6865-6930 \AA, 7580-7700 \AA). Then for each exposure we applied a weighted average between the cross-correlations of the single segments, 
where the weights applied to each segment are the inverse of its standard deviation (i.e., since we are in the photon-noise dominated regime, the higher the S/N, the greater the weight) and the depths of the lines in the model.
For a range of K$_{\rm p}$ values from 0 to 300 \kms, in steps of 1 \kms, we averaged the in-transit cross-correlation functions after shifting them in the planetary restframe. This is done by subtracting the planetary radial velocity calculated for each spectrum as $v_{p}=K_{\rm p} \times \sin{2\pi \phi}$, with $\phi$ the orbital phase. 
We thus created the K$_{\rm p}$ versus V$_{\rm sys}$ maps that are used to verify the real planetary origin of any possible signal.
We evaluated the noise by calculating the standard deviation of the K$_{\rm p}$ versus V$_{\rm sys}$ maps, where $|V|>70$ \kms, i.e., far from where any stellar or planetary signal is expected.
The significance of the detections (Table \ref{tab:crosscorr}) was calculated by dividing the best K$_{\rm p}$ cross-correlation function for the noise, and by fitting a Gaussian function to the result.

\begin{figure*}
\centering
\includegraphics[width=\linewidth]{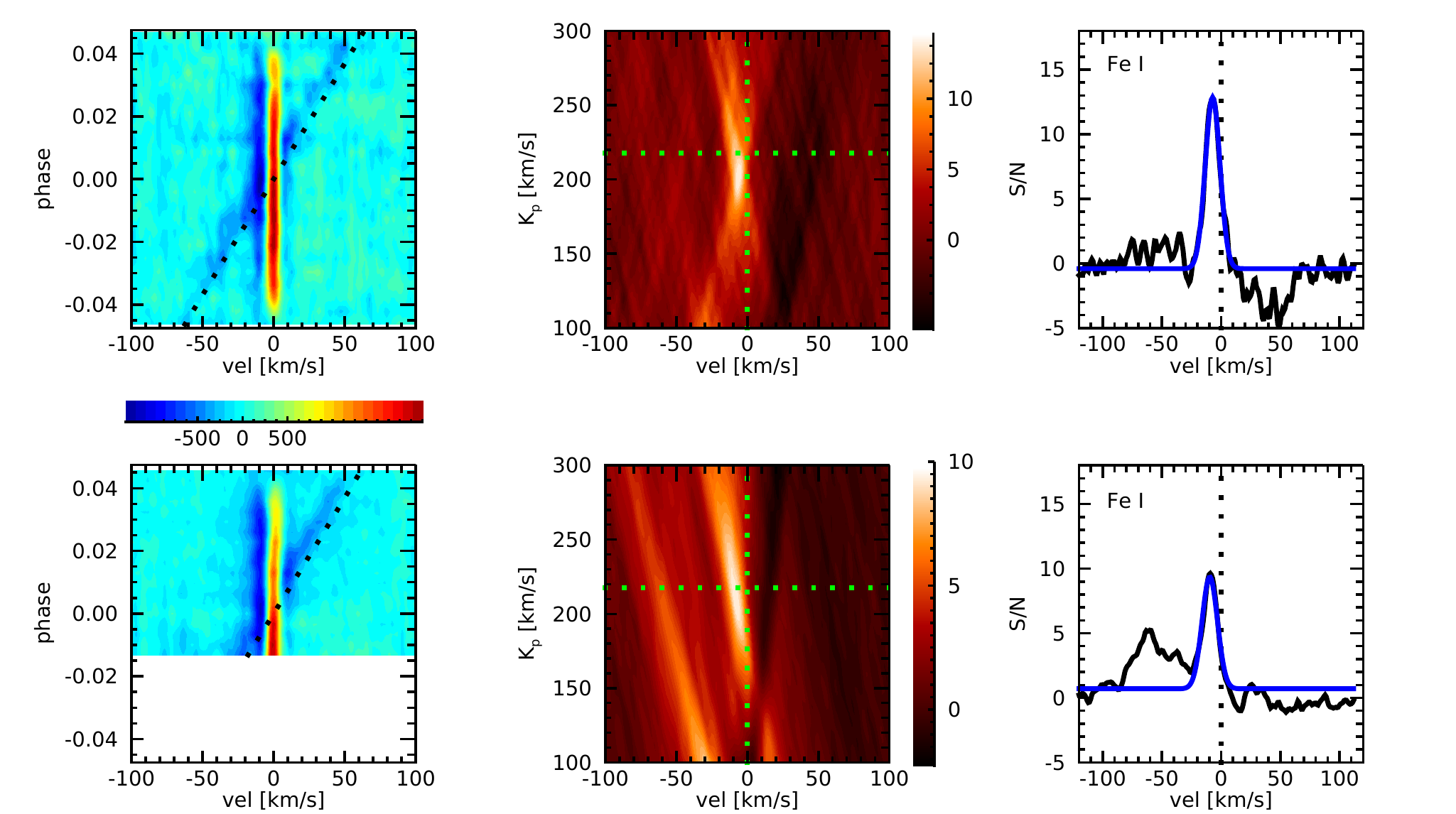}
\caption{Cross-correlation of the data with the \ion{Fe}{i} model. Shown are the results for the 1-UT transit (top), and those for the 4-UT transit (bottom).
The first column shows the contour plots of the temporal variation of the cross-correlation, with a dashed line to show the expected planetary velocity; the colorbar scale is in ppm.
The second column is the K$_{\rm p}$ vs velocity map after summing the cross-correlations for different K$_{\rm p}$, with colorbar scale in S/N. Green dashed lines are centered on the expected planetary position. 
The third column shows the final S/N of the detection, with the performed Gaussian fit in blue, as calculated at the theoretical K$_{\rm p}$ value.}
\label{fig.Fe1}
\end{figure*}

\begin{figure*}
\centering
\includegraphics[width=\linewidth]{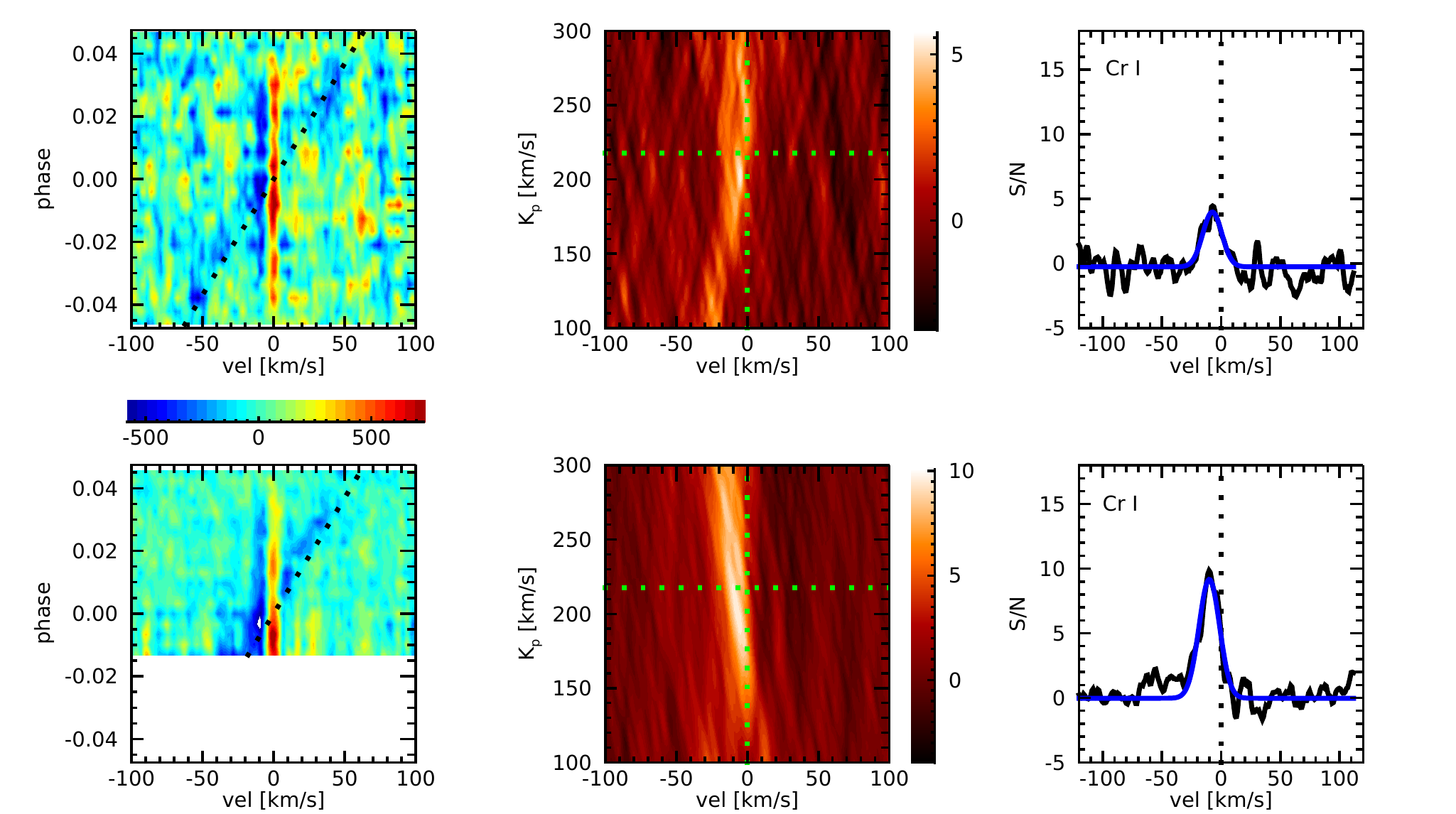}
\caption{Same as Fig. \ref{fig.Fe1}, but for \ion{Cr}{i}.}
\label{fig.Cr1}
\end{figure*}

\begin{figure*}
\centering
\includegraphics[width=\linewidth]{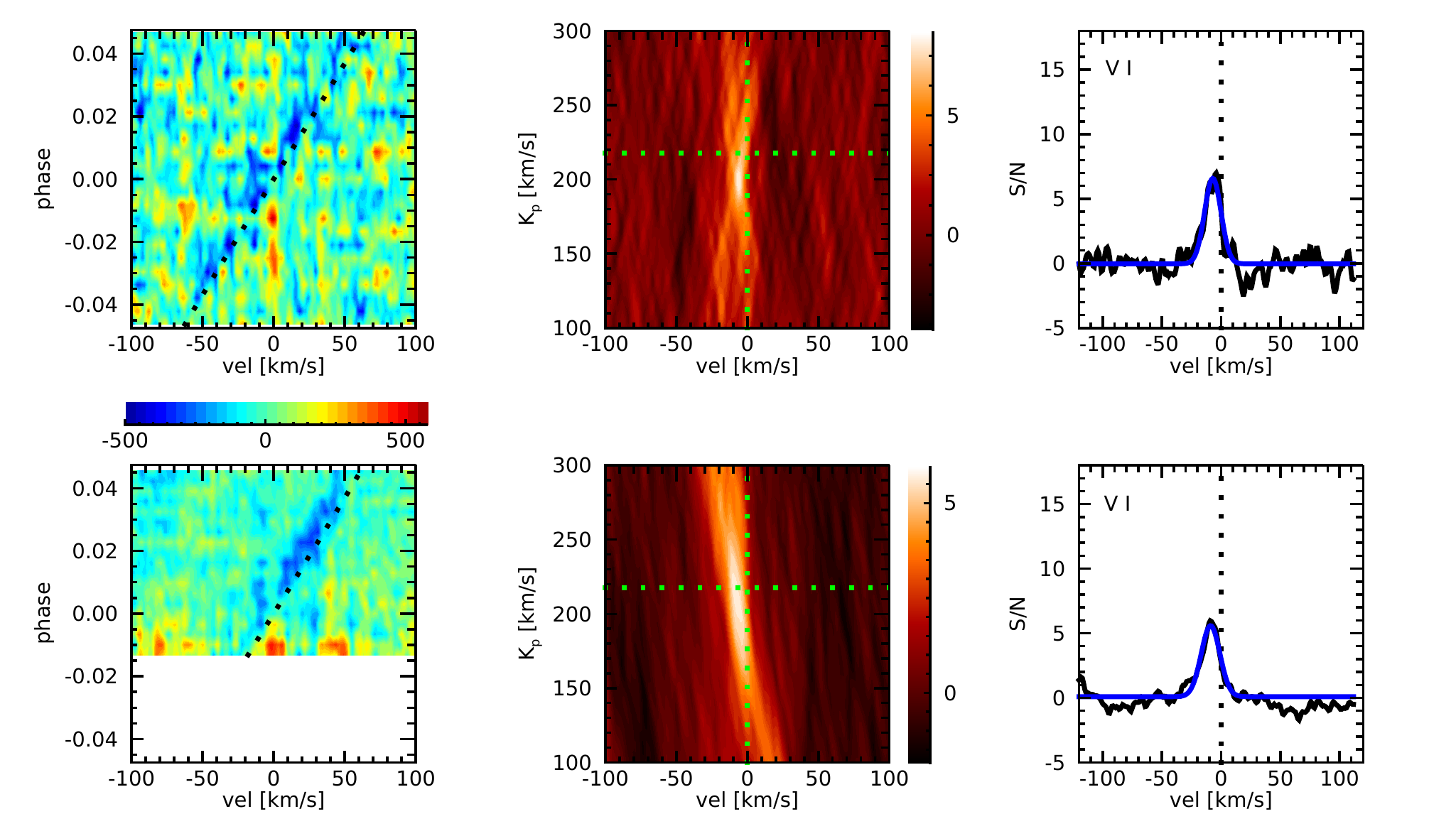}
\caption{Same as Fig. \ref{fig.Fe1}, but for \ion{V}{i}.}
\label{fig.V1}
\end{figure*}

\begin{figure*}
\centering
\includegraphics[width=\linewidth]{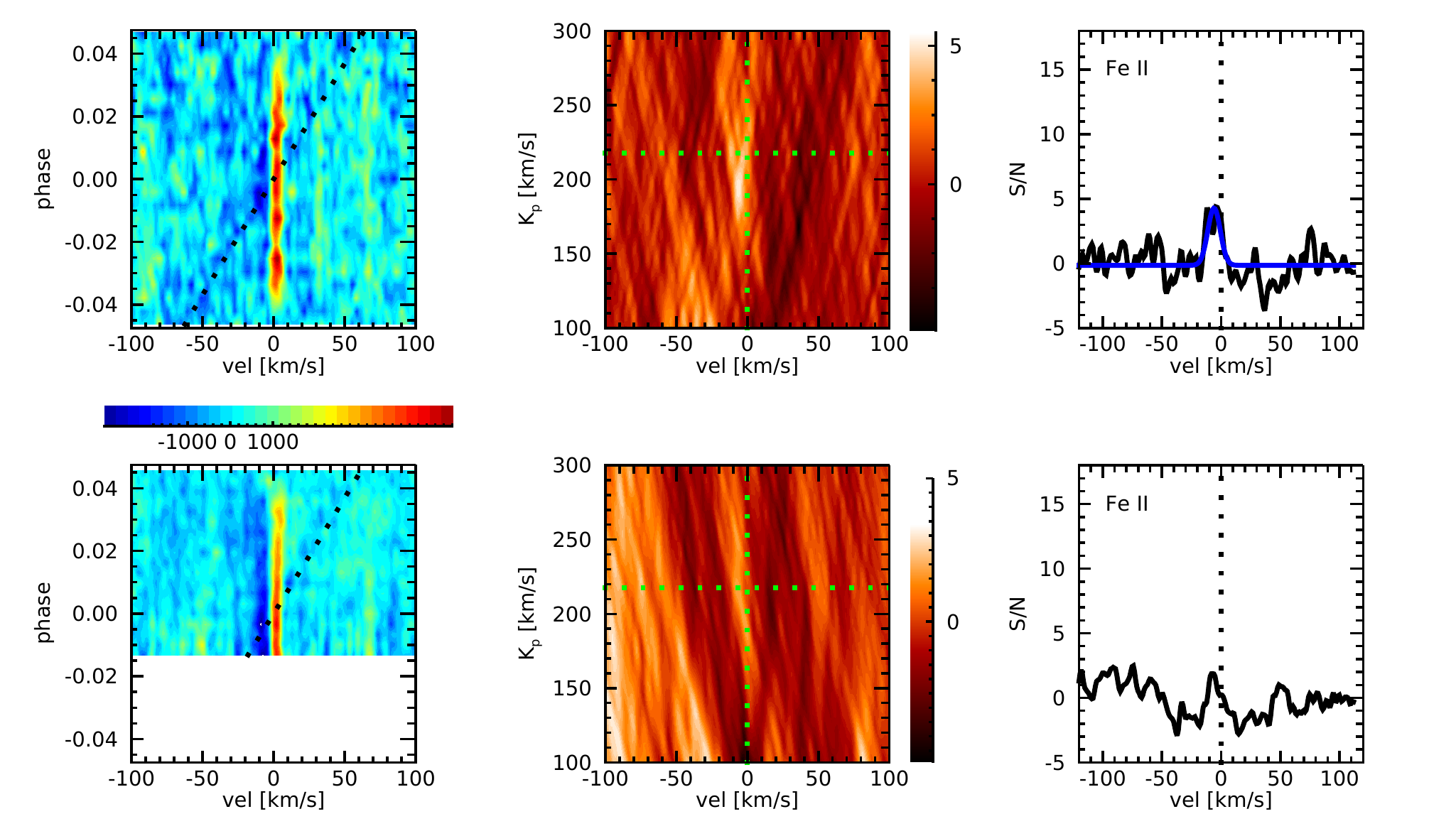}
\caption{Same as Fig. \ref{fig.Fe1}, but for \ion{Fe}{ii}. }
\label{fig.Fe2}
\end{figure*}

We confirm previous detections of \ion{Fe}{i} (Fig. \ref{fig.Fe1}), \ion{Cr}{i} (Fig. \ref{fig.Cr1}), and \ion{V}{i} (Fig. \ref{fig.V1}). We also confirm the presence of \ion{Fe}{ii} in the planetary atmosphere (Fig. \ref{fig.Fe2}), which has been previously debated \citep{2020arXiv200605995B,2020arXiv200611308H}. A net blueshift is present for all the detections.
We find no evidence of the presence of \ion{Ti}{i} (66 ppm at 3$\sigma$), \ion{Ti}{ii} (460 ppm), VO (18 ppm), TiO (12 ppm), CaH (36 ppm), or ZrO (20 ppm), given the accuracy of the linelists used.

In Table \ref{tab:crosscorr} we present a summary of the detections in the two transits.
The best K$_{\rm p}$ is often lower than the theoretical value (K$_{\rm p}\sim$218 \kms). This is possibly due to the atmospheric blueshift variability discussed in Sect. \ref{sec:planetaryCCF}.
It is interesting to note that while for \ion{Fe}{i}, \ion{Fe}{ii,} and \ion{V}{i} the detection significance (calculated at the theoretical K$_{\rm p}$) is higher for the complete 1-UT transit (with no detection of \ion{Fe}{ii} with the 4-UT), it is the opposite for the case of \ion{Cr}{i}. 
We recall that the two datasets were acquired with different resolving power (R$\sim$140,000 for   1-UT versus R$\sim$70,000 for   4-UT). We  also note that the efficiency of the two observing modes is different, and in particular the transits were observed before the fiber upgrade \citep{espresso2020}.
It is hard in this case to make a direct comparison of what is better between resolving power and S/N when looking for species by using the cross-correlation technique since the two datasets are not compatible due to the incompleteness of the 4-UT transit. It is  probable, however, that a combination of the two variables is a key factor, depending also on which wavelength region  the lines of the species that are searched for fall into.

\section{Discussion and conclusions\label{sec:discussion}}

We observed two transits of the UHJ WASP-121b with ESPRESSO. 
By analyzing the strange shape of the in-transit RVs, we were able to prove that this is a case of atmospheric RM, showing the presence of (mostly) Fe in the planetary atmosphere and its blueshift by using only the stellar RVs. 
The atmospheric RM is confirmed to be a valid way to detect in the atmosphere of exoplanets lines that are present in the stellar mask used to calculate the RVs.
By studying the exoplanet transmission spectrum, we more than doubled the number of previously detected atomic species in the atmosphere of WASP-121b, reinforcing previous detections of Na D1, Na D2, and H$\alpha$, and adding Li, Mg, K, and \ion{Ca}{ii} H\&K (and possibly H$\beta$ and Mn).
Lithium detection, despite its low amplitude ($\sim$0.2\%), is significant at $>6\sigma$ level. This is a remarkable result, considering that it is achieved with only one partial transit with the 4-UT mode and we were able to confirm it   in the planetary restframe with the 2D tomography. This became possible thanks also to the fact that there is no lithium line in the stellar spectrum, thus we are working at the continuum level with a very high S/N. Lithium was first claimed in an exoplanetary
atmosphere by \citet{2018A&A...616A.145C} for WASP-127b, but with low-resolution transmission spectroscopy \citep[and not confirmed with high resolution,][]{allart20}. \citet{tabernero_esp} and this work present the first detections of Li in planetary atmospheres at high resolution, also providing  2D tomographic evidence of their coherence with the planetary restframe. The detection of a trace element like Li in exoplanet atmospheres is an important step, as it can help in the understanding
of planet formation history and lithium depletion in planet-hosting stars \citep[e.g.,][]{2008A&A...489L..53B,2009Natur.462..189I,2018A&A...616A.145C}.
We note that we cannot detect lithium in the 1-UT transit because of the  insufficient S/N.

\begin{figure}
\centering
\includegraphics[width=\linewidth]{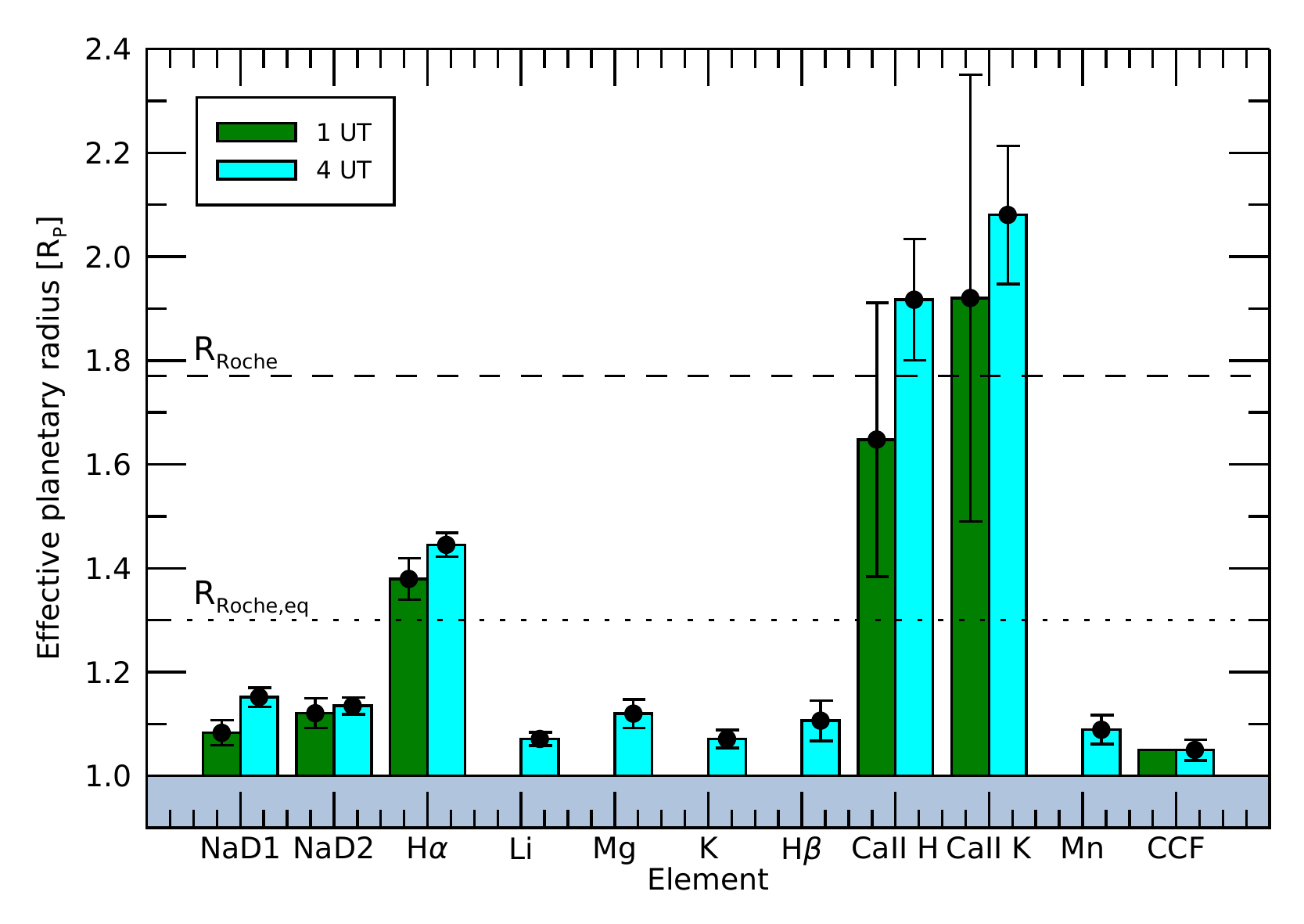}
\caption{Effective planetary radius for each element detected, shown for both the transits analyzed. Error bars refer to 1$\sigma$. The horizontal dashed and dotted lines represent respectively the Roche lobe radius of 1.77 $\rp$ and the transit-equivalent Roche lobe radius for the transiting exoplanet case of $\sim$1.3 $\rp$.}
\label{fig.atmoheight}
\end{figure}

In Fig. \ref{fig.atmoheight} we show the effective planet radius for the detected elements, which is calculated assuming $R^2_{\rm eff}/ R^2_{\rm p}=(\delta+h)/\delta$, where $\delta$ is the transit depth (from Table \ref{tabParameters}) and $h$ is the line contrast \citep[e.g.,][]{2020A&A...635A.171C}.
As expected, features like \ion{Ca}{ii} and H$\alpha$ are present up to very high altitudes in the atmosphere, as in the case of other hot exoplanets \citep[e.g.,][]{2018NatAs...2..714Y,2019A&A...632A..69Y,2019A&A...628A...9C}. 
The absorptions from \ion{Ca}{ii} H\&K lines are very prominent, probing atmospheric layers close to and possibly beyond the planetary Roche radius (R$_{\rm Roche} \sim 1.77$ $\rp$). 
As the Roche lobe is elongated toward the star, in a transiting configuration it extends to about 2/3 of the extension to the L1 Lagrange point \citep{2008ApJ...676L..57V}, with an equivalent value of the transiting Roche lobe of $\sim$1.3 $\rp$ for WASP-121b \citep[][]{sing}. Ionized species were already detected at these high altitudes in the atmosphere of WASP-121b with {\it HST} \citep[\ion{Fe}{ii} and \ion{Mg}{ii},][]{sing}. 
When considering the equivalent Roche lobe radius, our detections of \ion{Ca}{ii} H\&K and also H$\alpha$ are significantly beyond it at the 5.3$\sigma$, 5.9$\sigma$, and 6.2$\sigma$ level, respectively, for the case of the 4-UT transit. 
We note that the model spectra we used for the CLV+RM effect correction could be imperfect in the core of chromospheric lines such as \ion{Ca}{ii} H\&K \citep[e.g.,][]{2020arXiv200210595C}. However, the presence of strong absorption in the planet atmosphere is certain as it is also confirmed by the 2D tomographic map (Fig. \ref{fig.tomo_all}).
The slight asymmetries observed on these line profiles (Fig. \ref{fig.Ca_lines}) could be due to the extension beyond the Roche lobe radius, and thus could be probing an ongoing planetary atmospheric escape. 

\begin{figure}
\centering
\includegraphics[width=\linewidth]{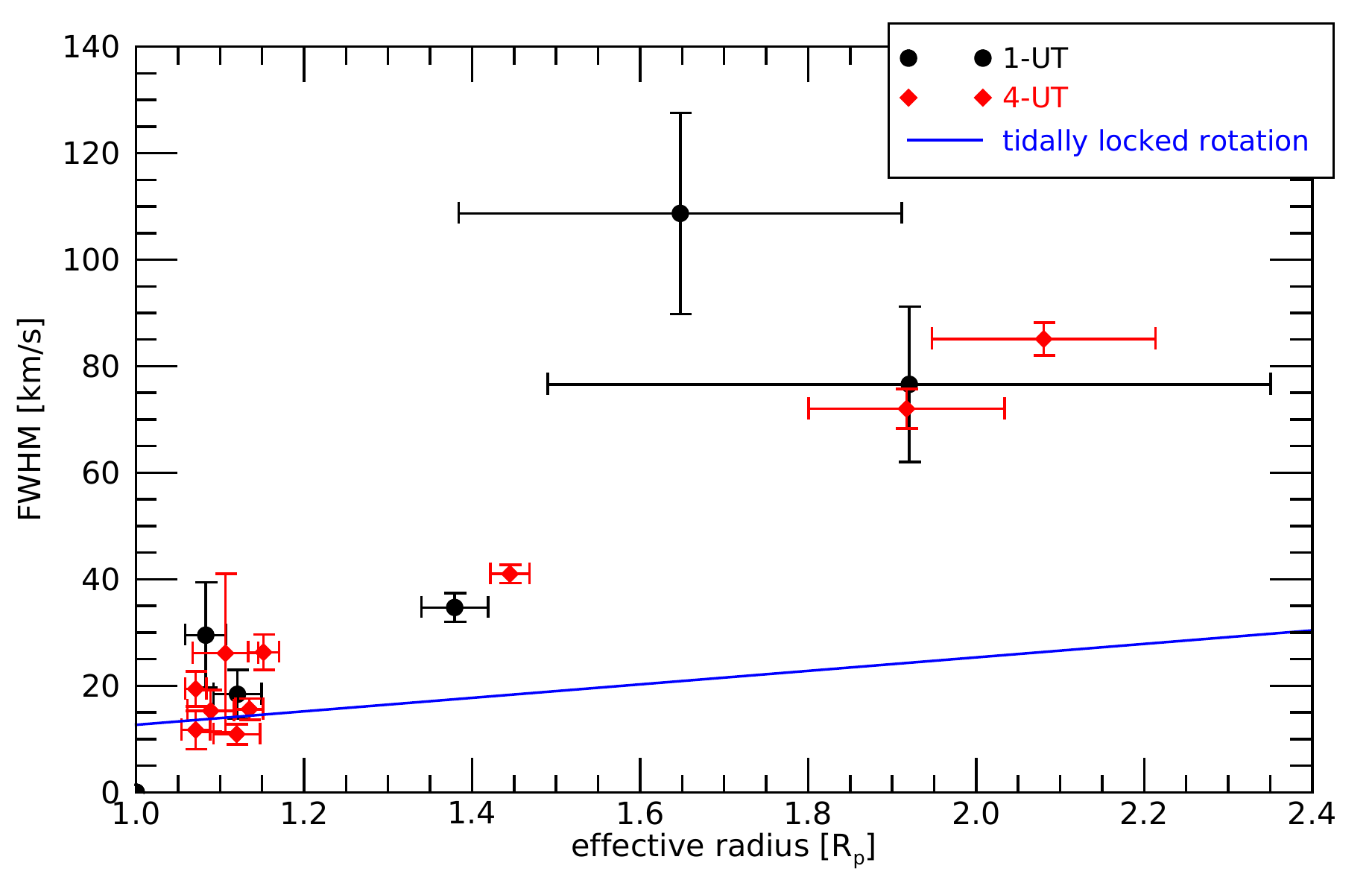}
\caption{Effective planetary radius vs FWHMs of the elements detected. The blue line shows the expected tidally locked planet's rotational broadening.}
\label{fig.rotation}
\end{figure}

Our detections probe different layers of the atmosphere, thus we compared the retrieved line widths to those predicted for a tidally locked WASP-121b (Fig. \ref{fig.rotation}). While for the low atmosphere tidally locked rotation can account for most of the thermal broadening \citep[also pointed out in][]{bourrier}, the full width at half maximum (FWHM) values of the deepest absorption lines H$\alpha$ and \ion{Ca}{ii} are significantly larger than those expected for a tidally locked rotating atmosphere. Tidally locked rotation is thus not the main source of atmospheric broadening for the higher layers of the atmosphere. In addition,  the blueshift of \ion{Ca}{ii} H\&K is larger than for the other detected lines.
The planet is probably experiencing atmospheric evaporation, thus the lines could be broadened by the expanding atmosphere via vertical winds or high-altitude jets.

We note that the line contrast of the detected atmospheric features is systematically (but not significantly) higher for the 4-UT data with respect to the 1-UT (Table \ref{tabdetections}, Fig. \ref{fig.atmoheight}). This is hardly due to the different spectral resolution, as in this case we would expect an opposite behavior (i.e., lines becoming deeper with higher resolving power). 
One possible cause is the atmospheric blueshift variability during transit shown in Sect. \ref{sec:planetaryCCF}. This in fact causes a spread of the lines when we sum all the single transmission spectra of the 1-UT transit. On the contrary, this does not happen for the 4-UT data, as they are restricted to the second part of the full transit where the atmospheric restframe is almost constant (Fig. \ref{fig.atmotrack}, left panel).
One more possible interesting explanation could be the different level of stellar activity during the two transits, with the star being more active when we observe deeper planetary atmospheric lines. 
The contrast difference  is seen for lines that are sensitive to stellar activity such as Na D, H$\alpha$, and \ion{Ca}{ii} H\&K.
We do not know exactly the timescales necessary for the atmosphere to react to stellar activity changes, but the temporal distance of $>$1 month between the two observed transits could be more than enough.

With one partial transit with the 4-UT MR mode we could detect more elements than with one complete transit with the 1-UT HR mode. This is due to the very high S/N of the 4-UT data, which exploits the light coming from a 16m-equivalent telescope despite the lower resolving power (R$\sim$70,000 versus R $\sim$138,000). In this case the S/N overcomes the spectral resolution when looking for single lines of atomic species.

In conclusion, ESPRESSO is able to deeply investigate the atmospheres of exoplanets and temporally resolve the atmospheric behavior; it will play a major role in the coming years in the field of exoplanet characterization. Here we showed its potential applied to the atmospheric analysis of WASP-121b, which is confirmed to be one of the most intriguing UHJs and that definitely deserves future follow-up investigations.

%

\begin{acknowledgements}
We thank the referee for their useful comments that helped improving the clarity of the manuscript.
The authors acknowledge the ESPRESSO project team for its effort and dedication in building the ESPRESSO instrument.
FB acknowledges financial support from INAF through the ASI-INAF contract 2015-019-R0, and M. Rainer for helpful discussions on the Fourier transform of the CCF. 
This work has been carried out within the framework of the National Centre of Competence in Research PlanetS supported by the Swiss National Science Foundation. R. A. acknowledge the financial support of the SNSF.
This work was supported by FCT - Funda\c{c}\~{a}o para a Ci\^{e}ncia e a Tecnologia through national funds and by FEDER through COMPETE2020 - Programa Operacional Competitividade e Internacionaliza\c{c}\~{a}o by these grants: UID/FIS/04434/2019; UIDB/04434/2020; UIDP/04434/2020; PTDC/FIS-AST/32113/2017 \& POCI-01-0145-FEDER-032113; PTDC/FIS-AST/28953/2017 \& POCI-01-0145-FEDER-028953; PTDC/FIS-AST/28987/2017 \& POCI-01-0145-FEDER-028987.
O.D.S.D. is supported in the form of work contract (DL 57/2016/CP1364/CT0004) funded by FCT.
This project has received funding from the European Research Council (ERC) under the European Union's Horizon 2020 research and innovation programme (project Four Aces grant agreement No 724427).

This work has made use of data from the European Space Agency (ESA) mission {\it Gaia} 
(\url{https://www.cosmos.esa.int/gaia}), processed by the {\it Gaia} 
Data Processing and Analysis Consortium (DPAC, 
\url{https://www.cosmos.esa.int/web/gaia/dpac/consortium}). Funding for 
the DPAC has been provided by national institutions, in particular the 
institutions participating in the {\it Gaia} Multilateral Agreement.

\end{acknowledgements}

%

\end{document}